\documentclass[twocolumn,showpacs,preprintnumbers,amsmath,amssymb]{revtex4}

\usepackage{graphicx}
\usepackage{dcolumn}
\usepackage{bm}

\def\bea{\begin{eqnarray}}
\def\eea{\end{eqnarray}}

\def\pp{\mbox{$p$-$p$} }
\def\auau{\mbox{Au-Au} }

\def\pbpb{\mbox{Pb-Pb} }
\def\aa{\mbox{A-A} }
\def\nn{\mbox{N-N} }

\def\pt{$p_t$ }

\def\deta{$\eta_\Delta$ }
\def\dphi{$\phi_\Delta$ }

\def\erf{\text{erf} }

\begin{document} 

\preprint{Version 2.6}


\title{

Azimuth multipoles  vs minimum-bias jets in 2D angular correlations on $\eta$ and $\phi$
}

\author{Thomas A. Trainor}
\address{CENPA 354290, University of Washington, Seattle, WA 98195}


\date{\today}

\begin{abstract}
Angular correlations measured in \pp and heavy ion collisions at the Relativistic Heavy Ion Collider (RHIC) include a same-side (SS) 2D peak. In peripheral \aa and p-p collisions the SS peak properties are consistent with predicted minimum-bias jet correlations. However, in more-central \auau collisions the SS peak becomes elongated on pseudorapidity $\eta$. 
Arguments have been proposed to explain the  SS peak $\eta$ elongation  in terms of possibly-fluctuating initial-state geometry multipoles coupled with radial flow to produce final-state momentum-space multipoles. Such arguments are based on Fourier decomposition of 2D angular correlations projected onto 1D azimuth.
In this analysis we show that measured correlation structure on $\eta$ (large curvatures) establishes a clear distinction between the SS 2D (jet) peak and 1D multipoles. Measured 2D peak systematics can predict inferred 1D Fourier amplitudes interpreted as ``higher harmonic flows.''  But 1D Fourier amplitudes alone cannot describe 2D angular correlations. The SS 2D peak remains a unique structure which can be interpreted in terms of parton scattering and fragmentation in all cases.

\end{abstract}


\maketitle

 \section{Introduction}


Data from nuclear collisions at the Relativistic Heavy Ion Collider (RHIC)  have been interpreted within a  hydrodynamic (hydro) context to demonstrate the formation of a thermalized, flowing partonic medium with small viscosity~\cite{qgp1,qgp2}. However, alternative analysis of spectrum and correlation data reveals a minimum-bias jet or {\em minijet} contribution whose variation with \aa centrality and collision energy seems to conflict with hydro expectations~\cite{axialci,daugherity,hardspec,fragevo,jetspec}.  A large-amplitude 2D peak at the origin in angular correlations on pseudorapidity $\eta$ and azimuth $\phi$ expected as a jet manifestation persists even in central \auau collisions, albeit the 2D peak is elongated on $\eta$ relative to a nominally symmetric jet cone~\cite{axialci,daugherity}. 

The competition between minimum-bias jets and flows to explain the final-state structure of RHIC collisions  has recently intensified and now focuses on the SS 2D peak in $p_t$-integral angular correlations. The minijet interpretation of the SS peak in more-central \aa collisions~\cite{axialci,daugherity} has been challenged. 
The peak has  been reinterpreted in terms of ``triangular flow'' resulting from conjectured initial-state transverse geometry fluctuations (i.e., sextupole and other azimuth multipoles) coupled to radial expansion~\cite{gunther,luzum,sorensen}, or as resulting from interaction of initial-state  ``glasma flux tubes'' coupled to radial flow~\cite{mg,gmm}.  Interpretations based on initial-state \aa overlap geometry (the subject of this article) typically invoke a Fourier series to describe 2D angular correlations projected onto 1D azimuth. The Fourier sinusoids are interpreted to represent  ``higher harmonic flows''~\cite{luzum,alice}.

However, the 1D projection onto azimuth which is central to such arguments discards critical information contained in the pseudorapidity structure. A discrete Fourier series can describe any structure projected onto periodic azimuth.  A Gaussian peak narrow on azimuth must contribute significantly to several 1D Fourier terms.  But the SS 2D peak contribution should (and can) be distinguished from other 2D structure based on its strong pseudorapidity variation (curvature). 

In this analysis we invoke measured systematics of the minimum-bias ($p_t$-integral) SS 2D peak for 200 GeV \auau collisions to demonstrate that recently-reported  $v_m\{2\}$ measurements correspond to properties of the SS peak.  From systematics of the peak properties we can predict any higher harmonic flow $v_m\{2\}$ (Fourier coefficient) as a function of centrality and {\em $\eta$ exclusion cuts} meant to remove ``nonflow'' from $v_m$ measurements. 

In \pp and more-peripheral \aa collisions the SS 2D peak conforms closely to expectations for minimum-bias jets~\cite{ua1,porter2,porter3,ppprd,daugherity}. Based on comparison of spectra, correlations and pQCD calculations we conclude that the SS peak continues to represent jet production even in more-central \auau collisions~\cite{hardspec,fragevo,daugherity,jetspec}. 
The most likely interpretation of the SS peak mechanism in more-central \aa collisions remains parton scattering and fragmentation, with modification of fragmentation including $\eta$ elongation (polarization) of lower-momentum jet fragments. Thus, jet production is the mechanism behind the multipoles recently interpreted as higher harmonic flows.

This article is arranged as follows: 
We review analysis methods and possible correlation mechanisms in Secs.~\ref{analysis} through~\ref{isgeom}. We summarize measured 2D angular correlations from 200 GeV \auau collisions and the systematics of  minijet structure and the nonjet azimuth quadrupole in Sec.~\ref{angcorrdata}. We discuss Fourier series analysis and possible confusion between minijets and a nonjet quadrupole arising from some analysis methods in Secs.~\ref{periodpeak} and \ref{miniquad}. 
We then review conjectured initial-state geometry structure and related measures (triangular flow, higher harmonics) in Secs.~\ref{triangle} and \ref{luzumsec}.   In Sec.~\ref{predict} we present quantitative relations between minijet structure and azimuth multipoles inferred from various analysis techniques, with a direct comparison between higher multipoles predicted from 200 GeV minijet systematics and a recent measurement of higher harmonic flows from the LHC.

 \section{Analysis methods} \label{analysis}

We briefly introduce correlation analysis methods applied to nuclear collisions at the RHIC. Method details are described in Refs.~\cite{porter2,porter3,inverse,axialci,daugherity,davidhq,davidhq2,davidaustin}. 
Topics include \aa collision geometry, correlation measures and 2D correlation spaces. 
A detailed discussion of initial-state \aa geometry is presented in Sec.~\ref{isgeom}, a two-component angular correlations data model is presented in Sec.~\ref{angcorrdata} and Fourier series analysis relevant to 1D azimuth correlations is reviewed in Sec.~\ref{periodpeak}. 

\subsection{Initial-state \aa geometry}

Initial-state (IS) \aa collision geometry is described by the Glauber model relating the \aa differential cross section to participant-nucleon number $N_{part}$ and \nn binary-collision number $N_{bin}$~\cite{powerlaw}. A derived participant-nucleon mean path length $\nu = 2 N_{bin} / N_{part}$ can also be defined. Through the measured \aa differential cross section on charged-hadron multiplicity $n_{ch}$ within some angular acceptance the Glauber model  parameters are related to observed $n_{ch}$.

Optical $\epsilon_{opt}$~\cite{davidhq} and Monte Carlo $\epsilon_{MC}$~\cite{phoboseps} eccentricities have been invoked  to model the IS \aa overlap eccentricity required for interpretation of the final-state (FS) azimuth quadrupole measured by $v_2$. 
$\epsilon_{opt}$ assumes a smooth matter distribution across nuclei whereas $\epsilon_{MC}$ assumes that point-like participant nucleons are the determining elements.
{\em A priori} support for $\epsilon_{opt}$ derives from a conjecture that the {\em nonjet} azimuth quadrupole emerges from interactions at small $x < 0.01$ where one might expect onset of a smooth, saturated glue system (e.g.\ Glasma)~\cite{gluequad}.
{\em A posteriori} support for $\epsilon_{opt}$ is suggested by a simple systematic trend observed for the nonjet quadrupole (Sec.~\ref{2dquad}).

\subsection{Two-particle correlation measures} \label{corrmeas}

Two-particle correlations are structures in pair-density distributions on six-dimensional momentum space $(p_{t1},\eta_1,\phi_1,p_{t2},\eta_2,\phi_2)$. We visualize correlation structure in 2D subspaces $(p_t,p_t)$ and $(\eta_\Delta,\phi_\Delta)$ (defined below) which retain almost all structure within a limited $\eta$ acceptance such as the STAR Time Projection Chamber (TPC).
Correlations can be measured with {\em per-particle} statistic  $\Delta \rho / \sqrt{\rho_{ref}} = \rho_0\, (\langle r \rangle-1)$, where $\Delta \rho= \rho_{sib} - \rho_{ref}$ is the correlated-pair density, $\rho_{sib}$ is the sibling (same-event) pair density, $\rho_{ref}$ is the reference- or mixed-pair density, $\langle  r \rangle$ is the mean sibling/mixed pair-number ratio, and prefactor  $\rho_0 = \bar n_{ch} / \Delta \eta\, \Delta \phi  \approx d^2n_{ch} / d\eta d\phi$ is the charged-particle 2D angular density averaged over angular acceptance $(\Delta \eta,\Delta \phi)$~\cite{axialci,axialcd}. Pair ratio $r$ is averaged over kinematic bins (e.g. multiplicity, $p_t$, vertex position). Factorization $\rho_{ref} \approx \rho_0^2$ is assumed. 

The {\em per-particle} quadrupole component of 2D angular correlations resulting from some analysis method is  $A_Q\{\text{method}\} $ defined in Eq.~(\ref{method}). Correlations can also be measured with  {\em per-pair} statistic $\Delta \rho / \rho_{ref}$, including total azimuth quadrupole component $v_2^2\{2\}$ and higher multipoles $v_m^2\{2\}$. 
Variation of per-pair correlation measures with \aa centrality is typically dominated by a trivial $1/n_{ch}$ trend, or in the case of $v_m$ a $1/\sqrt{n_{ch}}$ trend.

\subsection{Number correlations on $\bf (p_t,p_t)$ or $\bf (y_t,y_t)$}

2D correlations on $p_t$ or transverse rapidity $y_t = \ln[(p_t + m_t) / m_\pi]$ ($m_\pi$ is assumed for unidentified hadrons) are complementary to 4D angular correlations in 6D two-particle momentum space. 
Correlations on angle differences $(\eta_\Delta,\phi_\Delta)$ and momenta $(p_t,p_t)$ [or rapidity $(y_t,y_t)$] can be defined for like-sign (LS) and unlike-sign (US) charge combinations and also for same-side (SS) and away-side (AS) azimuth subregions of angular correlations (defined below). Manifestations of different correlation mechanisms (e.g.\ so-called soft and hard components, Sec.~\ref{hadrprod}) can be clearly distinguished in the four combinations of charge-pair type and azimuth subspace, with distinctive forms for each of the LS and US charge combinations and for SS and AS azimuth subspaces~\cite{porter2,porter3}. Any conjectured correlation mechanism must accommodate all such observed systematic trends.

\subsection{Number and $\bf p_t$ angular correlations on $\bf (\eta_\Delta,\phi_\Delta)$}

Angular correlations can be formed by integrating over the entire $(p_t,p_t)$ pair acceptance (minimum-bias angular correlations) or over subregions~\cite{porter2,porter3}. Examples of the latter include  ``trigger-associated'' dihadron correlations resulting from asymmetric cuts on $(p_t,p_t)$~\cite{dihadron}.

Two-particle angular correlations are defined on 4D momentum subspace $(\eta_1,\eta_2,\phi_1,\phi_2)$. Within acceptance intervals where correlation structure is approximately invariant on mean polar or azimuth angle (e.g.\ $\eta_\Sigma = \eta_1 + \eta_2$) angular correlations can be {\em projected by averaging} onto difference variables (e.g.\ $\eta_\Delta = \eta_1 - \eta_2$) without loss of information to form {\em angular autocorrelations}~\cite{axialcd,inverse}. 2D subspace ($\eta_\Delta,\phi_\Delta$) is then visualized. 
Symbol $\Delta x$ denotes the detector acceptance on parameter $x$. 
The pair angular acceptance on azimuth can be separated into a same-side (SS) region ($|\phi_\Delta| < \pi / 2$) and an away-side (AS) region ($|\phi_\Delta| > \pi / 2$). The SS region includes {\em intra}\,jet correlations (hadron pairs within single jets), while the AS region includes {\em inter}\,jet correlations (hadron pairs from back-to-back jet pairs).

Attempts to isolate 2D or 1D correlation structure from a (large) combinatoric background generally follow one of two methods: (a) model fits to 2D histograms~\cite{axialci,daugherity} and (b) ZYAM subtraction from 1D dihadron correlations on azimuth~\cite{tzyam}. 
2D angular correlations on $(\eta_\Delta,\phi_\Delta)$ are observed to include a few elements accurately described by simple functional forms, as described in Sec~\ref{angcorrdata}. Two of the elements have been interpreted in terms of jet correlations~\cite{axialci,jetspec}.

\section{Hadron production models} \label{hadrprod}

Several classes of hadron production models are invoked to describe high-energy nuclear collisions at the RHIC and LHC. One class consists of \nn superposition models, the limiting case being Glauber {\em linear} superposition (GLS). 
A second class is based on the limiting case of a homogeneous bulk medium produced through substantial parton and/or hadron rescattering with significant equilibration of low-energy partons. High-energy partons are expected to probe thermalized-medium properties via jet structure modification. A third class incorporates ``nonsmooth'' initial conditions plus hydrodynamic evolution to produce novel correlation structure in the final state.

\subsection{N-N superposition models}

Hadron production from \nn collisions, as in unmodified single collisions (linear), or collisions modified by the \aa environment (nonlinear), is superposed according to the Glauber model to describe hadron production in the \aa final state. Hadron production in \nn collisions is described by a theoretical two-component model including longitudinal projectile fragmentation and transverse scattered parton fragmentation~\cite{lund,pythia,hijing}. Modified hadron production in \aa collisions~\cite{hardspec} can be described by alteration of the pQCD fragmentation (parton splitting) process~\cite{fragevo}.

Longitudinal nucleon fragmentation results from ``soft'' four-momentum transfers between projectiles. Transverse parton fragmentation results from ``hard'' momentum transfers between constituent partons at some momentum fractions $x_1$, $x_2$ leading to dijet production. The terms ``soft'' and ``hard'' refer to the IS four-momentum transfer, not to the momenta of FS hadrons. Correlations from the soft component play a negligible role in more-central \aa collisions, but the soft component remains the dominant single-particle hadron production mechanism. The nonjet azimuth quadrupole is introduced in Sec.~\ref{angcorrdata} as a ``third component,'' first observed in more-central \aa collisions.

\subsection{Bulk-medium formation and freezeout}

A second model class emerging from lower-energy heavy ion programs~\cite{radial1}  describes production of a thermalized bulk medium (hadronic and/or partonic) in which the dominant collision mechanism is hydrodynamic flows~\cite{hydro}. The signature manifestation in \aa angular correlations is an azimuth quadrupole described by function $\cos(2\phi_\Delta)$ and interpreted as elliptic flow~\cite{2004}.

Medium formation and thermalization rely on hadron~\cite{rqmd,urqmd} and possible parton rescattering. Hadrons emerge from the bulk medium through a freezeout process exhibiting radial~\cite{radial1,radial2} as well as elliptic~\cite{2004} flow. Hadrochemical trends and $p_t$ spectrum structure are interpreted to indicate a thermal system supporting flows~\cite{radial1,radial2}.

\subsection{Initial-state geometry plus radial flow}

More recently, higher multipoles in initial-state \aa geometry coupled to radial flow in a bulk medium have been proposed as a possible mechanism for some final-state correlation structure. For instance, the SS 2D peak is conjectured to be a consequence of initial-state geometry multipoles~\cite{gunther}, geometry fluctuations~\cite{sorensen} or Glasma flux tubes~\cite{gmm}. Theoretical models are represented by Monte Carlos such as AMPT~\cite{ampt} and NexSpheRIO~\cite{rio}. IS geometry models compete directly with \nn superposition models. That dichotomy is a subject of this article.

\section{Initial-state \aa Geometry} \label{isgeom}

Initial-state \aa overlap geometry was first modeled by smooth nuclear-matter densities to obtain ``optical'' eccentricity $\epsilon_{opt}$ as a function of centrality. Motivated in part by interest in alternative interpretations of angular correlation features possibly attributed to jets the Glauber Monte Carlo was extended to derive $\epsilon_{MC}$ from the distribution of point-like participant nucleons~\cite{phoboseps}. It is conjectured that IS geometry fluctuations coupled with radial flow may produce flow structure appearing as jet-like FS correlations~\cite{sorensen}. The role of conjectured fluctuations in IS phase-space geometry may be relevant to interpretation of such jet-like structure.

\subsection{IS transverse phase space}

IS structure is present in both momentum space and configuration space, whereas FS correlation structure is observed only in momentum space. IS momentum-space structure (e.g., scattered-parton distributions) may be transported with some fidelity to FS momentum space via a separately-measured process (e.g., parton fragmentation to jets), whereas IS configuration-space structure (e.g., conjectured IS geometry fluctuations) requires coupling to flows for manifestation in FS momentum space. 

Minimum-bias parton scattering is a common feature of IS momentum space in all high-energy nuclear collisions.  The scattered-parton distribution represents a large range of projectile-nucleon momentum fraction $x$, whereas the IS nonjet quadrupole may emerge only from low $x < 0.01$. Whether large-angle-scattered partons near mid-rapidity fragment to detectable FS jets in \aa collisions or whether they thermalize rapidly to drive radial flow is a major issue for RHIC collisions. 

\subsection{IS azimuth power spectrum}

The IS  \aa transverse geometry (at $x \approx 1$) may have three components: a static (``optical'') part at fixed impact parameter $b$,  a contribution from fluctuating $b$, both represented by even azimuth multipoles, and a conjectured stochastic part (point-like participant sampling) represented by a white-noise spectrum including even and odd multipoles.

\aa initial-state azimuth structure modeled at $x \approx 1$ by a Glauber Monte Carlo is described by a participant-nucleon autocorrelation~\cite{inverse}. For non-central \aa collisions the autocorrelation on azimuth difference $\phi_\Delta$ includes a few even-$m$ sinusoids dominated by $m=2$ and phase-correlated with impact parameter $b$, a uniform combinatoric background and a delta function $\propto N_{part}$ (self pairs) representing participant-nucleon sampling noise.
By the Wiener-Khinchine theorem the Fourier transform of the azimuth autocorrelation is a power spectrum represented by eccentricity elements $E_m^2 = N_{part}^2 \epsilon_m^2$, with per-pair eccentricity measures~\cite{gluequad}
\bea \label{optical}
\epsilon_{m,MC}^2 &=& \epsilon_{m,opt}^2 + \sigma^2_{\epsilon_m} + \delta\epsilon_{m}^2~~\text{$m$ even} \\ \nonumber
&=&  \delta\epsilon_{m}^2~~\text{$m$ odd}.
\eea
Eccentricity $\epsilon_{m,opt}^2$ represents the ``elliptical'' \aa overlap region for fixed $b$, and $\sigma^2_{\epsilon_m}$ represents the eccentricity variance due to event-wise fluctuations in $b$.  Monte Carlo random sampling generates a power spectrum $\delta\epsilon^2_m \propto 1/N_{part}$ approximately uniform on $m$ corresponding to the self-pair contribution $\approx N_{part} \delta(\phi_\Delta)$ in the azimuth autocorrelation. For a stochastic process there should be no phase relation between noise amplitudes $\delta\epsilon^2_m$ and impact parameter $b$. 
All higher $m$ are present in the IS Monte Carlo spectrum and {\em might} appear in the final state to some extent {\em if} Monte Carlo sampling at $x \approx 1$ were a legitimate model of IS geometry relevant to FS hadron production for $x \leq 0.01$  and $\eta = 0$.

\subsection{Azimuth power spectrum centrality trends}

\begin{figure}[h]
  \includegraphics[width=1.65in,height=1.65in]{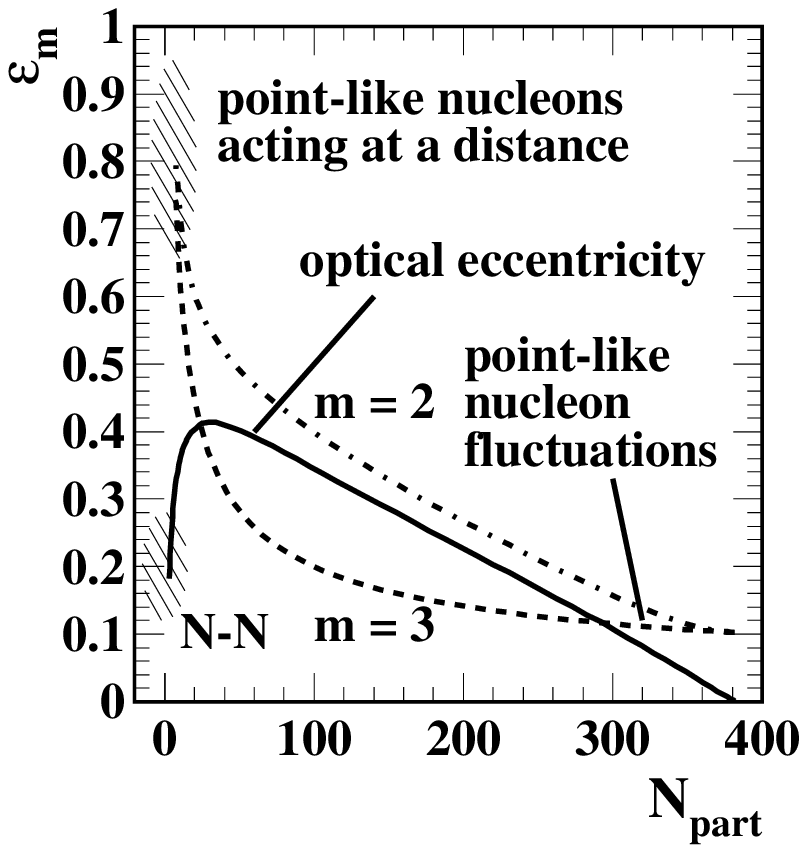}
  \includegraphics[width=1.65in,height=1.65in]{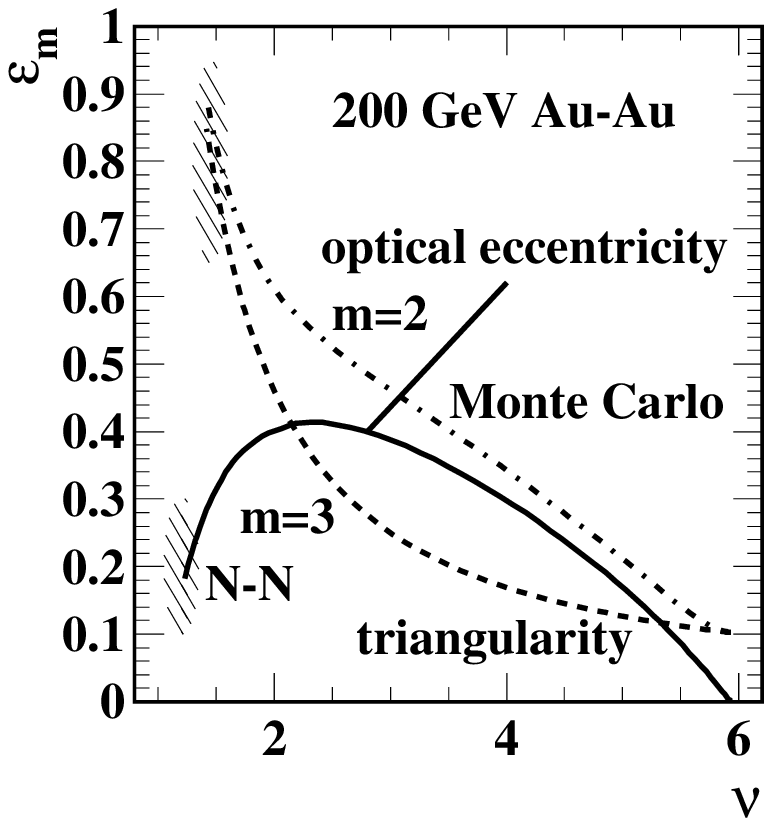}
\caption{\label{epps}
Centrality trends for optical and Monte Carlo Glauber initial-state azimuth multipoles with $m = 2$, 3, on participant-projectile-nucleon number (left panel) and binary \nn collisions $\nu$ per participant-nucleon  pair (right panel).
 } 
 \end{figure}

Figure~\ref{epps} shows centrality trends for $m = 2,$ 3 IS geometry power spectrum elements on participant-nucleon number $N_{part}$ (left panel) and mean participant pathlength $\nu$ (right panel). Plotted are optical eccentricity $\epsilon_{2,opt}$ (solid curve), Monte-Carlo eccentricity $\epsilon_{2,MC}$ (dash-dotted curve) and so-called ``triangularity'' $\delta \epsilon_3$ (dashed curve). 
From Eq.~(\ref{optical}) (and ignoring a possible $\sigma^2_{\epsilon_2 }$contribution) we have $\epsilon^2_{2,MC} = \epsilon^2_{2,opt} + \delta\epsilon^2_2$ with $\delta\epsilon^2_2 \approx 4/N_{part}$ and $\epsilon^2_{3,MC} = \delta\epsilon^2_3 \approx 4/N_{part}$. %
The optical eccentricity for 200 GeV \auau is parametrized by~\cite{davidhq}
\bea
\epsilon_{{2,opt}} = \frac{1}{5.4} \left[\log_{10}\left(\frac{3\, N_{bin}}{2}\right)\right]^{0.96}  \left[\log_{10}\left(\frac{1136}{N_{bin}}\right)\right]^{0.78}\hspace{-.23in}.
\eea

By comparing the trends in Fig.~\ref{epps} with published Monte Carlo results (e.g., Ref.~\cite{gunther} -- Fig. 2) we find that the stochastic contribution to the power spectrum for $m = 2,$ 3 is consistent with $O(1)\times 2/N_{part}$, with $O(1) \sim 0.5$-2 as expected for a per-pair fluctuation measure and Poisson statistics. Whether point-like sampling represents IS geometry with significant manifestations in FS correlation structure is an open question.

 \section{Two-component 2D data model} \label{angcorrdata}

Correlation data in the form of 2D angular autocorrelations~\cite{inverse} can be obtained for various $p_t$ cut configurations, including so-called ``trigger-associated'' dihadron correlations~\cite{dihadron} and minimum-bias ($p_t$-integral) correlations. For simplicity of illustration we restrict to $p_t$-integral data from 200 GeV \auau collisions~\cite{daugherity}. This exercise illustrates construction of a {\em necessary and sufficient} (N-S) 2D mathematical data model based entirely on data phenomenology (see App.~\ref{modeling}). The three model elements are subsequently interpreted physically in terms of longitudinal projectile fragmentation (soft component), transverse parton fragmentation (minijets) and  a nonjet azimuth quadrupole whose physical origin is in question.

\subsection{Angular correlation model function} \label{modelfunc}

2D data histograms on $(\eta_\Delta,\phi_\Delta)$ are not generally factorizable. However, minimum-bias histogram data obtained from RHIC \pp and \aa collisions can be represented as the sum of a few factored terms
\bea \label{histo}
h(\eta_\Delta,\phi_\Delta) = \sum_n f_n(\eta_\Delta)\, g_n(\phi_\Delta).
\eea
For minimum-bias data the series accurately represents all information in the data histogram with a few simple model functions. For some terms one factor may be approximately constant, further simplifying the model. 

Minimum-bias data from \auau collisions are described by three main elements: (a) a same-side (SS) 2D peak at the origin on $(\eta_\Delta,\phi_\Delta)$ well approximated by a 2D Gaussian for all minimum-bias data, (b) an away-side (AS) 1D peak on azimuth or ``ridge'' well approximated by AS azimuth dipole $[1 - \cos(\phi_\Delta)]/2$ for all minimum-bias data and uniform to a few percent on $\eta_\Delta$ (having negligible curvature), and (c) an azimuth quadrupole $\cos(2\phi_\Delta)$ also uniform on \deta to a few percent over the full angular acceptance of the STAR TPC. 
Model elements (a) and (b) together have been interpreted as minimum-bias jets or ``minijets''~\cite{fragevo}. Element (c), described as the (nonjet) azimuth quadrupole, is conventionally attributed to elliptic flow, a hydrodynamic phenomenon~\cite{2004} (but see App.~\ref{nonflowapp}). 

The 2D model function [Eq.~(\ref{histo}) equivalent] applicable to more-central \aa collisions is~\cite{axialci,daugherity,davidhq}
\bea \label{estructfit}
\frac{\Delta \rho}{\sqrt{\rho_{ref}}} \hspace{-.02in}  & = &  \hspace{-.02in}
A_0+  A_{2D} \, \exp \left\{- \frac{1}{2} \left[ \left( \frac{\phi_{\Delta}}{ \sigma_{\phi_{\Delta}}} \right)^2 \hspace{-.05in}  + \left( \frac{\eta_{\Delta}}{ \sigma_{\eta_{\Delta}}} \right)^2 \right] \right\} \nonumber \\
&+&  \hspace{-.02in} A_{D}\, \{1 +\cos(\phi_\Delta - \pi)\} / 2 + \hspace{-.02in}  A_{Q}\, 2\cos(2\, \phi_\Delta).
\eea
A 1D Gaussian on $\eta_\Delta$ modeling projectile nucleon fragmentation (soft component, negligible in more-central \auau collisions) and a 2D exponential modeling quantum correlations (HBT) and electron pairs (extremely narrow in more-central \auau collisions) are omitted for simplicity in discussion of more-central \aa collisions.
Quadrupole measure $A_Q$ is statistically consistent with jet measures $A_{2D}$ and $A_D$, permitting quantitative comparisons between jet and nonjet-quadrupole systematics.

 \begin{figure}[h]
  \includegraphics[width=1.65in,height=1.4in]{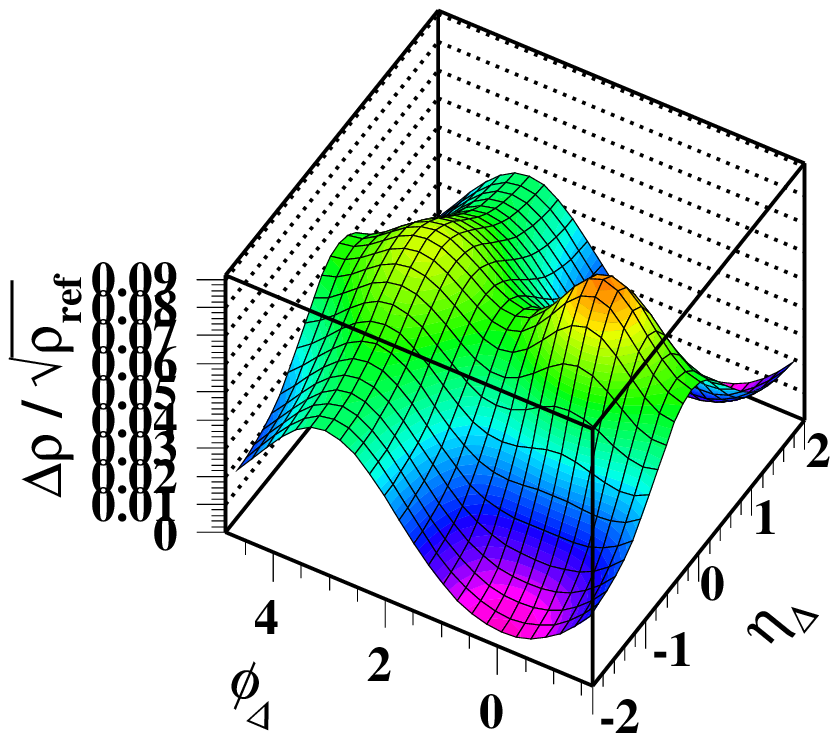}
\put(-100,80){\bf (a)}
  \includegraphics[width=1.65in,height=1.4in]{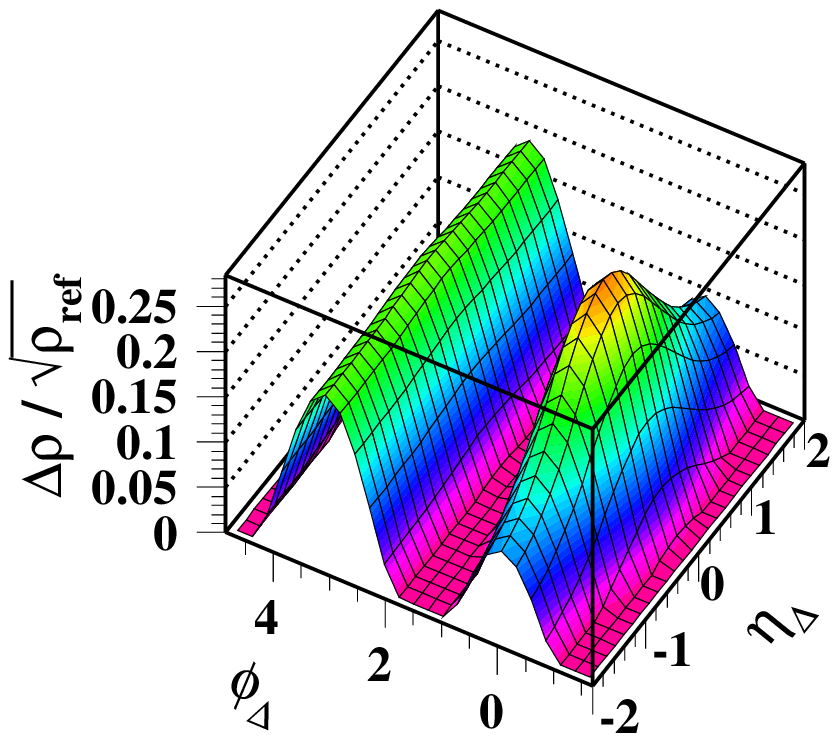}
 \put(-100,80){\bf (b)}\\
\includegraphics[width=1.65in,height=1.4in]{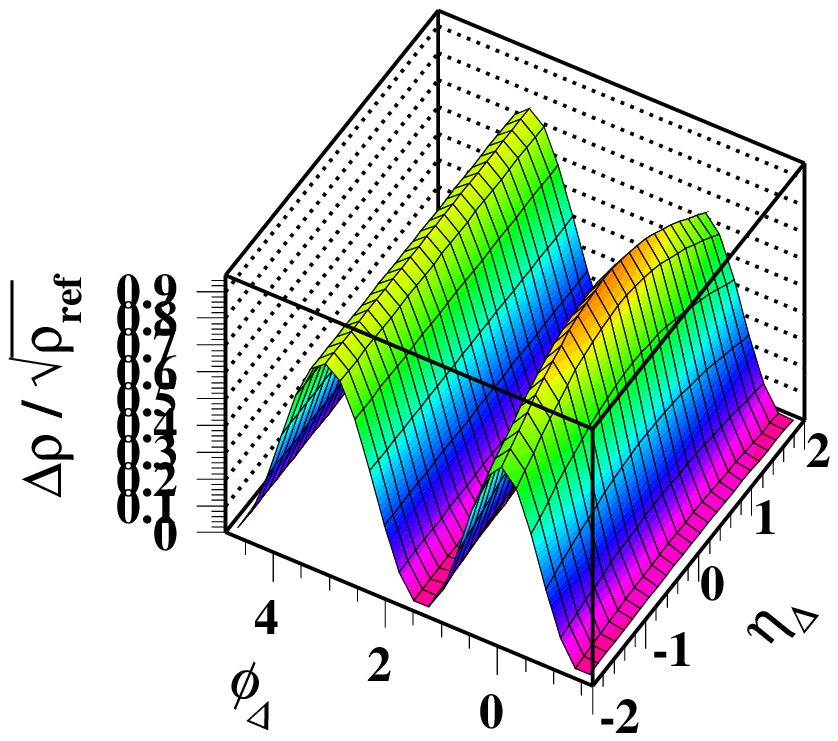}
\put(-100,80){\bf (c)}
  \includegraphics[width=1.65in,height=1.4in]{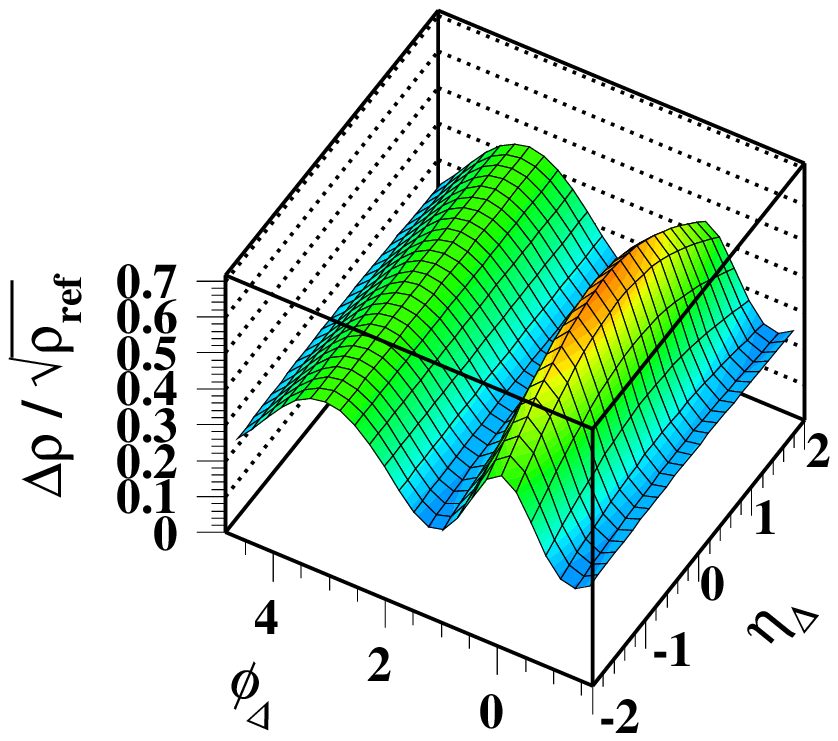}
\put(-100,80){\bf (d)}
\caption{\label{2dcorr} (Color online)
Angular correlation histograms for four centralities of 200 GeV \auau collisions based on fit parameters from Ref.~\cite{daugherity}. The centralities measured by $\nu$ are given by (a) 1.25 ($\sim$ N-N), (b) 2.5, (c) 4.5 and (d) 6 ($b = 0$).
 } 
 \end{figure}

Figure \ref{2dcorr} shows examples of 2D angular correlations from four centralities of 200 GeV \auau collisions based on fit parameters of Ref.~\cite{daugherity}. The centrality values $\nu$ correspond to (a) 1.25 ($\approx$ \nn collisions), (b) 2.5, (c) 4.5 and (d) 6 ($b = 0$). The histograms are plotted within the STAR TPC angular acceptance $|\eta_\Delta| < 2$ conventionally adopted for 2D correlation analysis.  2D correlation structure is accurately described by a simple mathematical model, and  2D fit residuals (r.m.s.\ amplitude) are typically less than 1\% of the SS 2D peak amplitude for more-central \auau collisions. Residuals for 1D projections onto azimuth are substantially smaller because of averaging.
We now consider the details of minijet and nonjet quadrupole systematics in turn.

\subsection{Final-state minijet systematics}

\begin{figure}[h]
  \includegraphics[width=1.65in,height=1.65in]{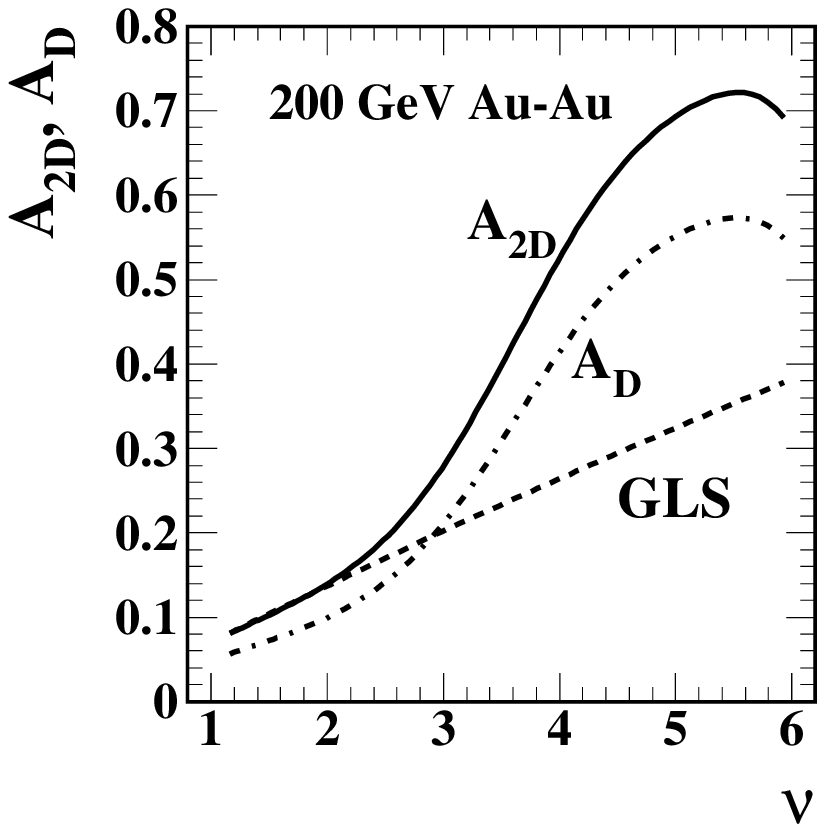}
  \includegraphics[width=1.65in,height=1.65in]{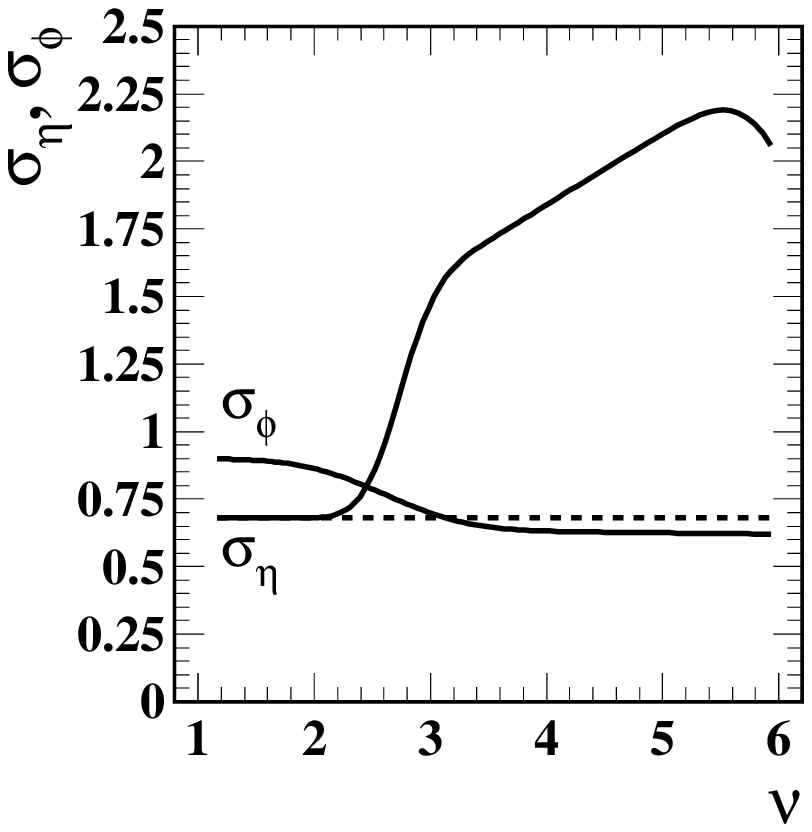}
\caption{\label{fitparams}
Left: Amplitude of the same-side 2D Gaussian $A_{2D}$ and away-side dipole $A_D$ fitted to minimum-bias 2D angular correlation data from 200 GeV \auau collisions~\cite{daugherity}.
Right: Fitted peak widths for the same-side 2D Gaussian. GLS indicates a Glauber linear superposition reference extrapolated from measured \pp collisions~\cite{ppprd}.
 } 
 \end{figure}

Figure \ref{fitparams} summarizes preliminary fitted SS 2D and AS 1D peak parameters vs centrality measure $\nu$ within the nominal STAR TPC angular acceptance $(\Delta \eta,\Delta \phi) = (2,2\pi)$~\cite{daugherity}. $A_{2D}$ in the left panel is the fitted amplitude of the SS 2D Gaussian function. Its two r.m.s.\ peak widths are shown in the right panel. AS dipole amplitude $A_D$ closely follows the SS 2D peak amplitude as expected for back-to-back dijets. There is smooth variation with centrality, but a ``sharp transition'' in SS 2D peak properties occurs near centrality $\nu \approx 3$.  Although the SS 2D peak becomes broad on \deta the peak curvature on $\eta_\Delta$ remains large in all cases. Large curvature and unique $p_t$ correlation structure differentiate the SS 2D peak from conjectured flow mechanisms.

The correlation parameters in Fig.~\ref{fitparams} were combined with a pQCD parton spectrum to predict equivalent jet fragment yields as hadron spectrum hard components~\cite{jetspec}. The predictions agree closely with previously-extracted spectrum hard components~\cite{hardspec}. In turn, differential spectrum hard components are described quantitatively by a full pQCD calculation~\cite{fragevo}. The agreement among pQCD theory, single-particle spectra and two-particle correlations within a common minijet framework provides strong support for a minijet interpretation of the SS 2D peak for all \auau centralities.

\subsection{Final-state nonjet quadrupole systematics} \label{2dquad}

\begin{figure}[h]
  \includegraphics[width=1.65in,height=1.65in]{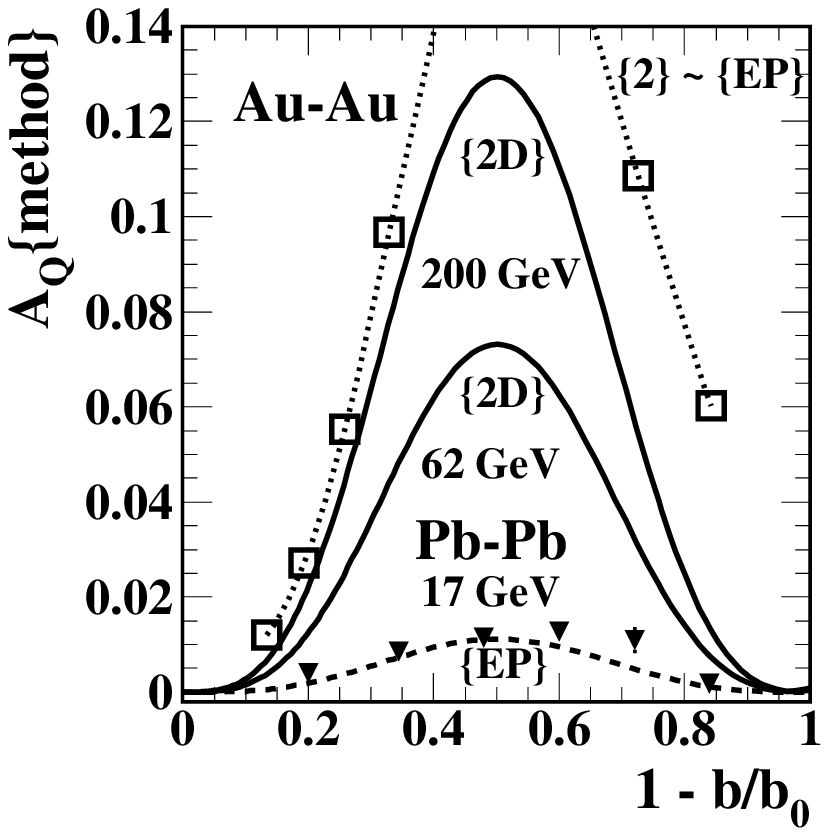}
  \includegraphics[width=1.65in,height=1.65in]{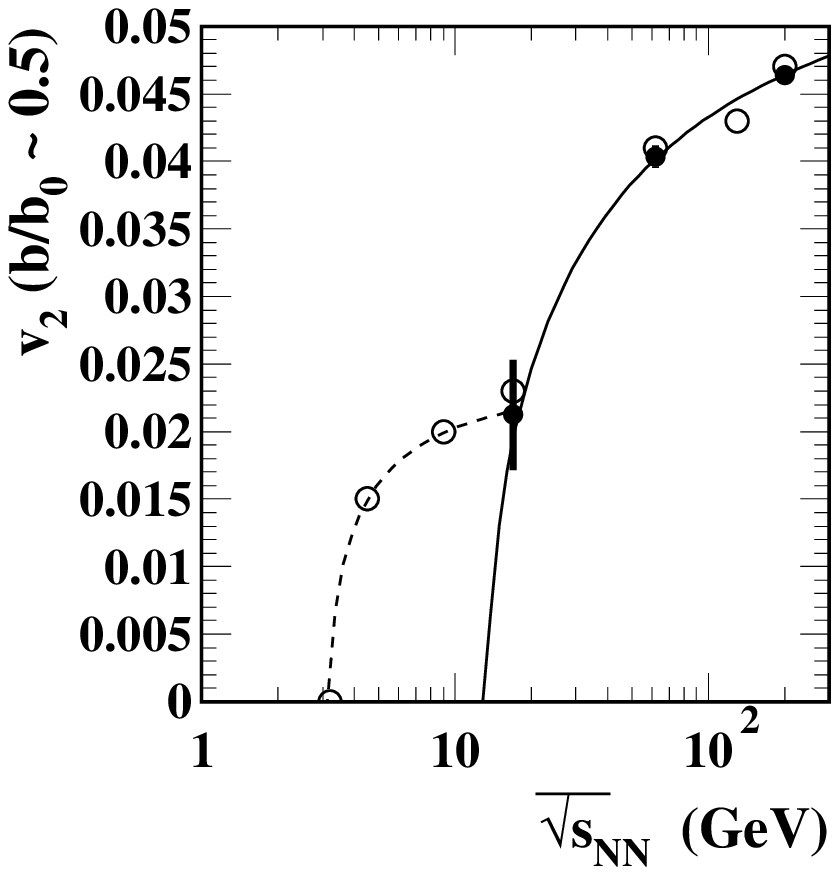}
\caption{\label{quads}
Left: Quadrupole amplitudes for different collision systems and analysis methods. Solid curves are $A_Q\{2D\}$ from Ref.~\cite{davidhq}. Open squares are transformed $v_2\{2\}$ data from Ref.~\cite{2004}. Solid triangles are similarly derived from $v_2\{EP\}$ data reported in Ref.~\cite{na49v2}. The curves are described in the text.
Right: Quadrupole data reported in the form of conventional elliptic flow measure $v_2$. The curves are appropriately transformed from the straight lines in Fig.~\ref{edep} (left panel).
 } 
 \end{figure}

Figure~\ref{quads} (left panel) shows centrality trends (solid curves) inferred from quadrupole data for 62 and 200 GeV \auau collisions reported in Ref.~\cite{davidhq}. Per-particle quadrupole amplitude $A_Q$ is related to conventional measure $v_2$ by
\bea \label{method}
A_Q\{\text{method}\}  = \rho_0(b) v_2^2\{\text{method}\} 
\eea
for $v_2$ methods \{2\} $\approx$ \{EP\} (nongraphical numerical methods) and \{2D\} (model fits to 2D histograms), where single-particle density  $\rho_0(b)$ is defined in Sec.~\ref{corrmeas}. Some of the 200 GeV $A_Q\{2\}$ data (open squares) fall above the upper plot boundary. $v_2\{EP\}$ data from 17 GeV \pbpb collisions are also included (solid triangles)~\cite{na49v2}. The same centrality trend (solid curves) describes the $A_Q\{2D\}$ data over a large range of energies.

\begin{figure}[h]
   \includegraphics[width=1.65in,height=1.65in]{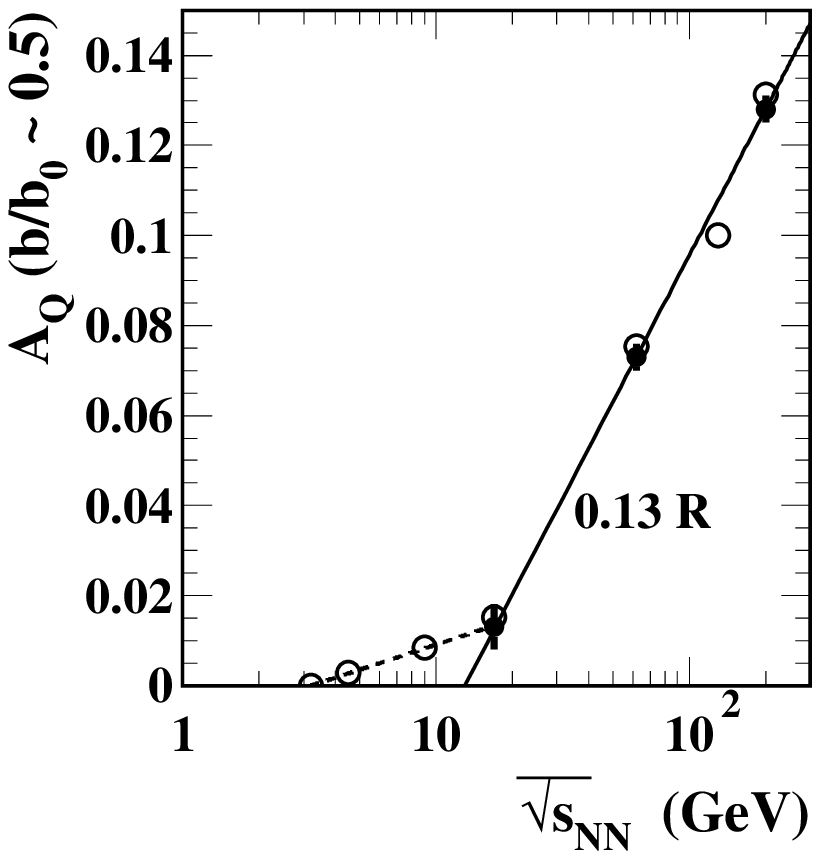}
 \includegraphics[width=1.65in,height=1.65in]{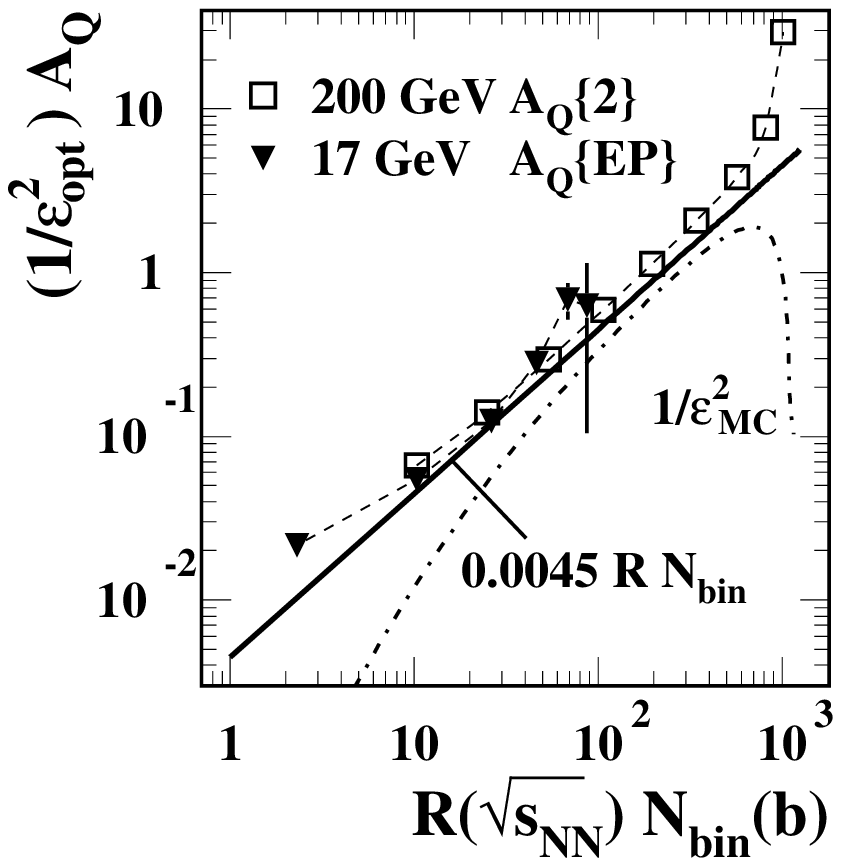}
\caption{\label{edep}
Left: Collision energy dependence of quadrupole amplitudes $A_Q$ for $v_2$ data from Ref.~\cite{2004} (open circles) and $A_Q\{2D\}$ reported in Ref.~\cite{davidhq}.  The solid line is from Eq.~(\ref{rroots}). The dashed line describes $v_2$ data in $A_Q$ format at lower energies~\cite{alicev2}.
Right: Universal systematics for nonjet quadrupole $A_Q\{2D\}$ reported in Ref.~\cite{davidhq} (bold solid line) compared to $A_Q\{2\}$ and $A_Q\{EP\}$ data presented in Fig.~\ref{quads} (left panel).
 }  
 \end{figure}

Figure~\ref{quads} (right panel) shows the energy systematics of quadrupole data in the form of conventional measure $v_2$, the square root of a per-pair correlation measure which tends to overemphasize small data values from systems with small particle multiplicities (e.g., lower collision energies). 
The open circles are derived from published $v_2\{EP\}$ data, the solid points from Ref.~\cite{davidhq}. Centrality choice $b/b_0 \approx 0.5$ ($\nu \approx 4.3$) minimizes the relative jet contribution to $v_2\{EP\}$. The solid and dashed curves are derived from the straight lines in Fig.~\ref{edep} (left panel) suitably transformed. 

Figure~\ref{edep} (left panel) shows the energy systematics of quadrupole data in the form of per-particle measure $A_Q$. The $A_Q$ trend above 13 GeV is described by the solid line $A_Q(\sqrt{s_{NN}};b/b_0 \approx 4.5) = 0.13\, R(\sqrt{s_{NN}})$~\cite{davidhq}, with
\bea \label{rroots}
R(\sqrt{s_{NN}}) = \ln\left(\sqrt{s_{NN}} / \text{13.5 GeV}\right).
\eea
The dashed line describing data below 13 GeV is $0.008 \ln\left(\sqrt{s_{NN}} / \text{3.2 GeV}\right)$ describing the well-known transition (sign change) from squeezeout (due to participant shadowing) to in-plane expansion.
That panel suggests qualitatively different quadrupole production mechanisms below and above 13 GeV, in contrast to Fig.~\ref{quads} (right panel) which might suggest continuation of Bevalac/AGS projectile-nucleon (nucleon cluster) collectivity to RHIC and LHC energies.

Figure~\ref{edep} (right panel) summarizes the measured centrality and energy dependence of {\em nonjet} quadrupole amplitude $A_Q\{2D\}$. All $p_t$-integral nonjet quadrupole data from \auau collisions are accurately summarized above 13 GeV by~\cite{davidhq}
\bea \label{davideq}
A_Q\{2D\} =  0.0045 R(\sqrt{s_{NN}}) \,N_{bin}\, \epsilon_{2,opt}^2
\eea
defining the bold solid line which transforms to the solid and dashed curves  in Fig.~\ref{quads} (left panel). The simple linear trend applies to $\epsilon_{opt}^2$, not $\epsilon_{MC}^2$ (dash-dotted curve). The centrality trend of $A_Q\{2D\}$ on $b / b_0$ (Fig.~\ref{quads} -- left panel) is Gaussian to good approximation, independent of collision energy above 13 GeV. The nonjet quadrupole centrality and $p_t$ dependence~\cite{davidhq,davidhq2} are independent of SS or AS jet structure. A unique characteristic of the nonjet quadrupole relative to the SS 2D peak is zero curvature on $\eta_\Delta$ within the STAR TPC $\eta$ acceptance, with important implications for IS geometry and the possible structure of flows on $\eta_\Delta$.

\section{Periodic peak arrays on azimuth} \label{periodpeak}

Because azimuth $\phi$ is a periodic variable {\em any} 1D structure on \dphi can be described {exactly} as a discrete Fourier cosine series
\bea \label{fourier1}
g(\phi_\Delta) &=& \sum_{m=0}  F_m\, \cos(m\phi_\Delta).
\eea
In principle, a factorized 2D structure $f(\eta_\Delta)g(\phi_\Delta)$ could be so expressed by multiplying Eq.~(\ref{fourier1}) through by the $\eta_\Delta$ factor (see App.~\ref{2dmult}). However, representing some 2D structure or its 1D projection by a few terms of a single 1D Fourier series can be misleading. 
In this section we consider the Fourier series representation of a periodic Gaussian peak array on 1D azimuth and projection of a SS 2D Gaussian onto azimuth. 
In the next section we discuss possible confusion between {\em jet-related} Fourier components of the SS 2D jet peak and the {\em nonjet} quadrupole (see Apps.~\ref{modeling} and \ref{nonflowapp}). 

\subsection{Fourier representation of 1D peak arrays}

Because 1D dihadron azimuth distributions are periodic, the peaks observed at $\phi_\Delta = 0$ (SS, same-side) and $\phi_\Delta = \pi$ (AS, away-side) are actually elements of separate periodic peak arrays described by cosine {series}. The SS array is centered on {even} multiples of $\pi$, the AS array on {odd} multiples. Nearest array elements outside a $2\pi$ interval (image peaks) produce significant structure within the observed interval and should be included in fit models. 

Each peak array (SS or AS) can be represented by a  Fourier series of the form
\bea \label{fourier}
S(\phi_\Delta;\sigma_{\phi_\Delta},n) &=&  \sum_{m=0}^\infty F_{m,n}\, \cos(m\,[\phi_\Delta - n\,\pi]),
\eea
where the $F_{m,n}$ are functions of r.m.s.\ peak width $\sigma_{\Delta \phi}$ defined below. 
Since $n$ is even for SS peak arrays ($+$) and odd for AS arrays ($-$) odd multipoles must be explicitly labeled as SS or AS. The terms represent $2m$ poles, e.g. dipole ($m=1$), quadrupole ($m=2$), sextupole ($m=3$) and octupole ($m = 4$), referring to cylindrical as opposed to spherical multipoles. 

 \begin{figure}[h]
  \includegraphics[width=1.65in,height=1.65in]{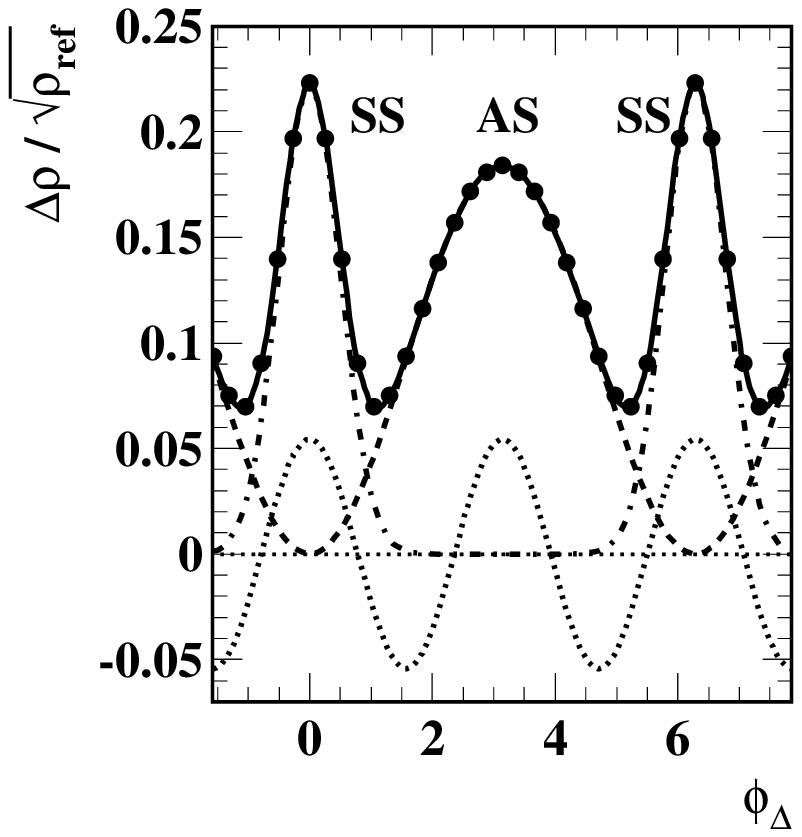}
  \includegraphics[width=1.65in,height=1.62in]{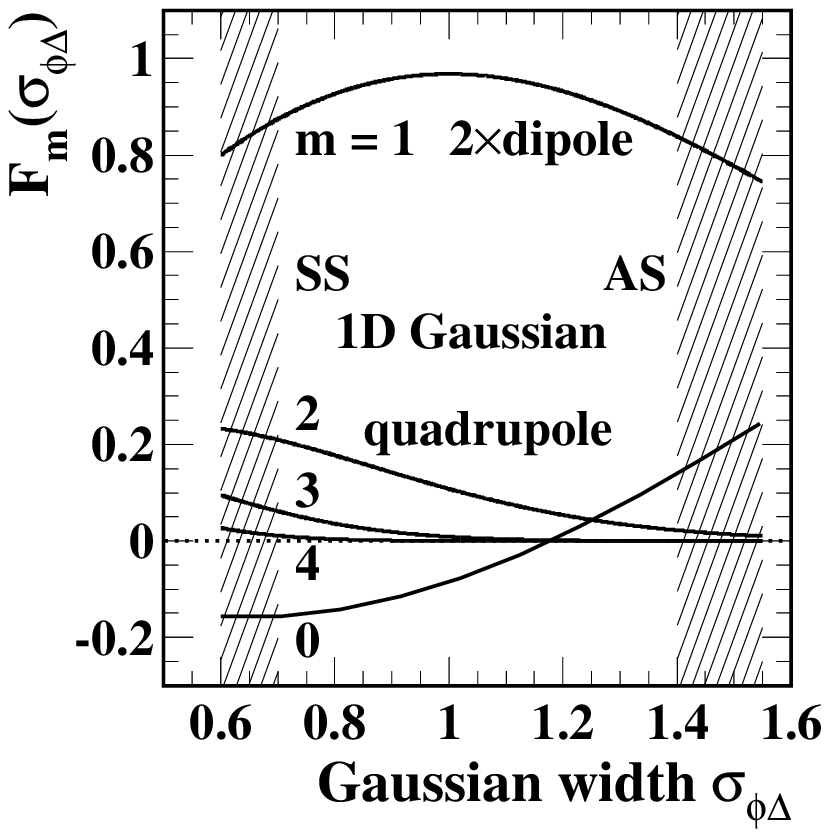}
\caption{\label{ortho}
Left: Periodic arrays of SS (dash-dotted) and AS (dashed) peaks. The SS peaks are Gaussians. The AS peaks are well-described by a dipole. The dotted sinusoid corresponds to the $m=2$ Fourier component of the SS peaks.
Right: Fourier amplitudes $F_m$ of a unit-amplitude Gaussian [Eq.~(\ref{fm})] vs peak width $\sigma_{\phi_\Delta}$.
 } 
 \end{figure}

Fig.~\ref{ortho} (left panel) illustrates the sum of peak arrays (solid points) for SS and AS peaks extending beyond one $2\pi$ period. The SS Gaussian peak array is the dash-dotted curve, the AS array with $\sigma_{\Delta \phi} \sim \pi / 2$ is the dashed curve (approximately pure dipole in this case). The dotted curve is the quadrupole term of the SS peak array, which would add a large ``nonflow'' contribution to $v^2_2\{2\}$ inferred from that distribution.

Fig.~\ref{ortho} (right panel) shows the Fourier amplitudes  $F_m$ of a unit-amplitude Gaussian array for the first five terms ($m \in [0,4]$) of Eq.~(\ref{fourier1}) as functions of the r.m.s.~peak width
\bea \label{fm}
F_m(\sigma_{\phi_\Delta}) &=& \sqrt{2/\pi}\, \sigma_{\phi_\Delta} \exp\left( - m^2 \sigma_{\phi_\Delta}^2 / 2\right).
\eea
As peak width $\sigma_{\phi_\Delta}$ increases, the number of significant terms in the series decreases. The limiting case is $\sigma_{\phi_\Delta} \sim \pi / 2$, for which the peak array is approximated by a constant plus dipole term $[1+\cos(\phi_\Delta)]/2$ (SS) or $[1 -\cos(\phi_\Delta)]/2$ (AS). 
For narrower (SS) peaks terms with $m > 1$ become significant, and  a Gaussian function is the more efficient representation. In particular, for $\sigma_{\phi_\Delta} \approx 0.65$ (typical for the SS jet peak) jet-related quadrupole amplitude $F_2 \approx 0.22$ represents the dominant jet-related nonflow contribution to $v_2^2\{2\} \sim v_2^2\{EP\}$ data. 

\subsection{Projecting a 2D Gaussian onto 1D azimuth} \label{projecteta}

The relation between an SS 2D peak projected to a narrow SS 1D Gaussian on azimuth and its Fourier components comes into play in Secs.~\ref{triangle} and \ref{luzumsec} which address recent claims of higher flow harmonics. 
The SS 2D peak, well-modeled by a 2D Gaussian, is distributed on $(\eta_\Delta,\phi_\Delta)$ within angular acceptance $(\Delta \eta,2\pi)$. We wish to determine the amplitude of the equivalent SS 1D Gaussian projected onto azimuth $\phi_\Delta$. We first consider the case that the entire $\eta$ acceptance $\Delta \eta$ is projected onto $\phi_\Delta$. We consider more-complex $\eta$ {\em exclusion} cuts in Sec.~\ref{predict} and App.~\ref{etacuts}. The projection factor is given by
\bea \label{gfac}
G(\sigma_{\eta_\Delta},\Delta \eta) \hspace{-.05in} &=&\hspace{-.05in} \frac{\int_0^{\Delta \eta} \hspace{-.03in }dx (\Delta \eta - x) \exp\left\{-x^2 / s^2 \right\}}{\int_0^{\Delta \eta} dx (\Delta \eta - x)} \\ \nonumber
&=&  \sqrt{\pi} \zeta\, \erf(1/\zeta) - \zeta^2 \left[ 1 - \exp(-1/\zeta^2) \right],
\eea
with $s = \sqrt{2} \sigma_{\eta_\Delta}$ and $\zeta = s / \Delta \eta$. $G(\sigma_{\eta_\Delta},\Delta \eta) \rightarrow 1$ as $\zeta \rightarrow \infty$. We consider an application with SS peak parameters for 0-5\% central \auau collisions and $\Delta \eta = 2$. 

 \begin{figure}[h]
  \includegraphics[width=1.65in,height=1.65in]{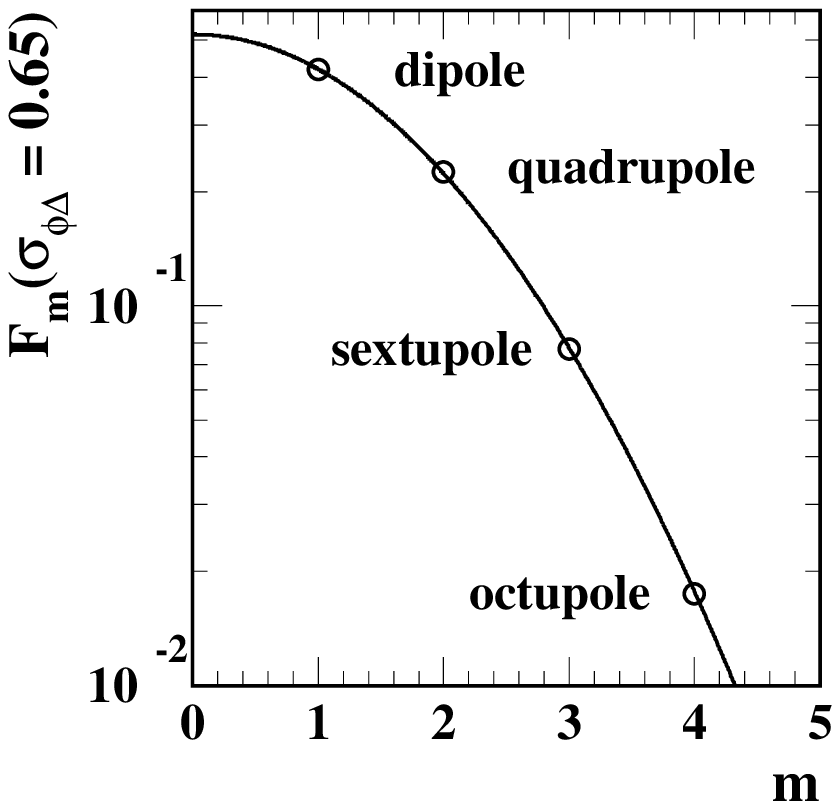}
  \includegraphics[width=1.65in,height=1.65in]{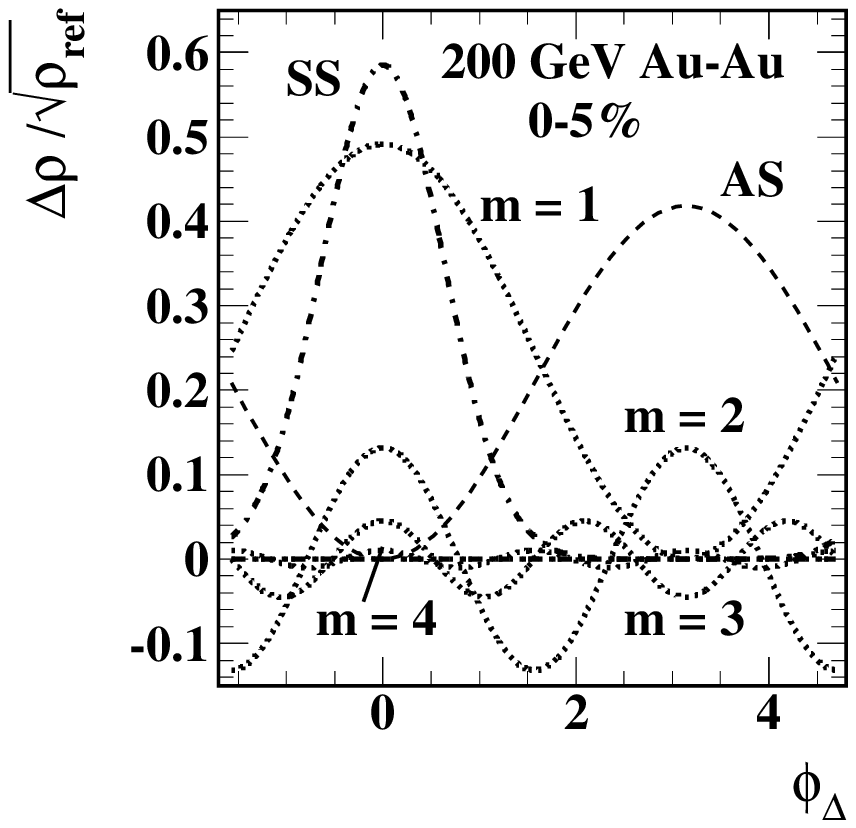}
\caption{\label{fourier}
Left: Evaluation of Eq.~(\ref{fm}) for four values of $m$.
Right: Typical results for 1D azimuth correlations from near-central \auau collisions. Same-side peak (dash-dotted curve) and broad away-side peak approximated by a dipole (dashed sinusoid). The narrow SS peak can be decomposed into several sinusoids ($m = 1...4$, dotted curves).
 } 
 \end{figure}

Fig.~\ref{fourier} (left panel) shows Eq.~(\ref{fm}) for $\sigma_{\phi_\Delta} = 0.65$, with the first few multipole coefficients marked for reference (open circles).
Figure~\ref{fourier} (right panel) illustrates 2D$\rightarrow$1D projection and Fourier decomposition of the SS peak. The SS (dash-dotted curve) and AS (dashed curve) peaks model jet correlations from 0-5\% central \auau collisions where the nonjet quadrupole amplitude is consistent with zero~\cite{daugherity,davidhq}. From Fig.~\ref{fitparams}  ($\nu \approx 6$) we obtain $A_{2D} \approx 0.7$, calculate $G(2.25,2) \approx 0.85$ from Eq.~(\ref{gfac}) and obtain the $F_m$ from Fig.~\ref{fourier} (left panel). The jet-related multipole amplitudes are then given by
\bea \label{aqss}
2 \rho_0(b)\, v^2_m\{SS\}(b) \hspace{-.05in} &=& \hspace{-.05in} F_m(\sigma_{\phi_\Delta})G(\sigma_{\eta_\Delta},\Delta \eta) A_{2D}(b). 
\eea
Figure~\ref{fourier} (right panel) shows the corresponding azimuth multipoles as dotted sinusoids with amplitudes from Eq.~(\ref{aqss}). Note that $A_Q\{SS\} = \rho_0(b)\, v^2_2\{SS\}(b)$ is the jet-related quadrupole amplitude, and $2A_Q\{SS\} = 0.225\times 0.85 \times 0.7  = 0.135$ defines the $m=2$ sinusoid.

\section{Minijets $\bf vs$ nonjet quadrupole} \label{miniquad}

The interplay between (mini)jet structure and nonjet quadrupole contributions to measured $v_2$ data plays a central role in the interpretation of RHIC data.  The nonjet quadrupole $A_Q\{2D\}$ is a unique phenomenon with centrality, energy and $p_t$ dependence distinct from minijets. Some $v_2$ analysis methods confuse jet and nonjet-quadrupole structure, leading to possible crosstalk between jets and quadrupole in $v_2$ data (see App.~\ref{nonflowapp}). We here focus on azimuth quadrupoles ($m = 2$) within a contiguous $\eta$ acceptance (e.g., $|\eta| < 1$), then consider higher multipoles and $\eta$ exclusion cuts  in subsequent sections.

\subsection{Jet-related vs nonjet quadrupole}

The azimuth quadrupole measured by $v_2$ is conventionally attributed to ``elliptic flow.'' Contributions to $v_2$ from possible nonhydro mechanisms are called ``nonflow.'' In the present context we refer instead to a nonjet quadrupole (what might be attributed to elliptic flow) and a jet-related quadrupole ($v_2$ contribution mainly from jets and mainly from the SS 2D jet peak). The distinction between flow and nonflow has been extensively discussed (e.g., Ref.~\cite{2004} and see App.~\ref{nonflowapp}).

\begin{figure}[h]
  \includegraphics[width=1.65in,height=1.6in]{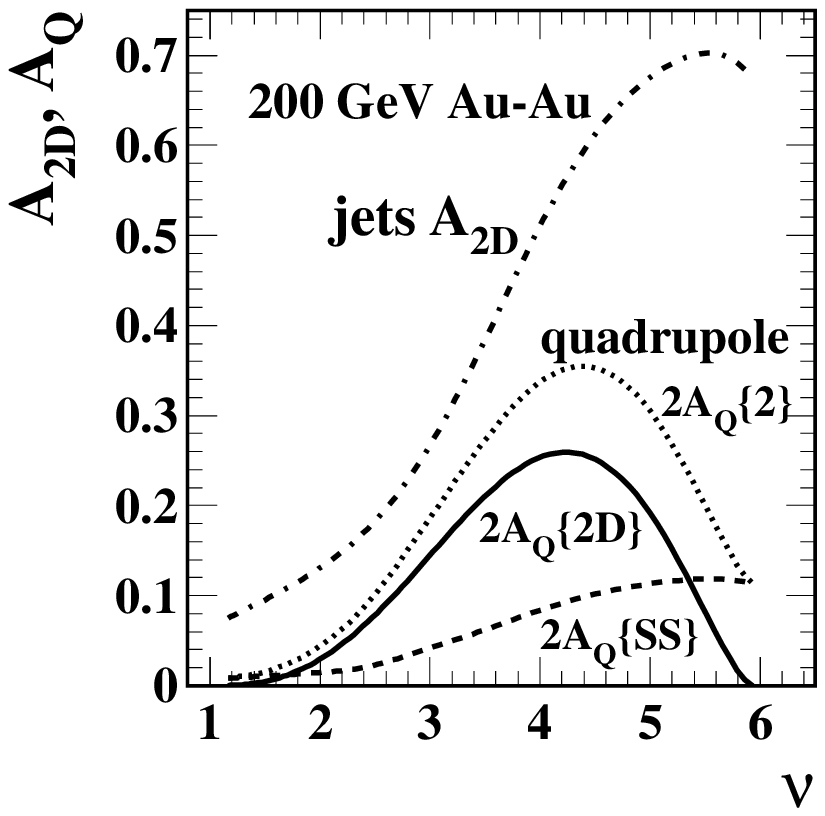}
  \includegraphics[width=1.65in,height=1.63in]{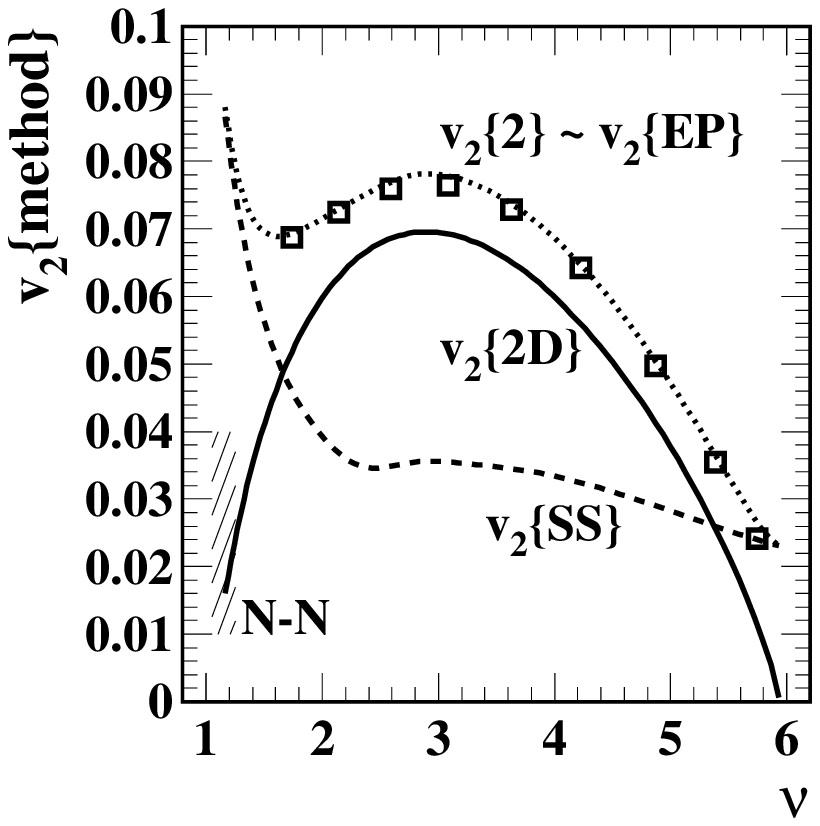}
\caption{\label{nonflow}
Left: SS 2D (jet) peak amplitude $A_{2D}$, ``nonflow'' SS peak quadrupole component $A_Q\{SS\}$ and nonjet quadrupole $A_{2D}$ amplitudes, with $A_Q\{2\} = A_Q\{2D\} + A_Q\{SS\}$~\cite{gluequad}. 
Right: Quadrupole amplitudes $A_Q$ converted to conventional measure $v_2$. Open squares are $v_2\{2\}$ data from~\cite{2004}.
 }  
 \end{figure}

Figure~\ref{nonflow} (left panel) shows the centrality dependence of SS 2D peak amplitude $A_{2D}$ and three quadrupole amplitudes related by  $A_Q\{2\} = A_Q\{2D\} + A_Q\{SS\}$~\cite{gluequad}. $A_Q\{2D\}$ is defined by Eq.~(\ref{davideq})~\cite{davidhq} and $A_Q\{SS\}$ by Eq.~(\ref{aqss}) using SS peak parameters from Ref.~\cite{daugherity} summarized in Fig.~\ref{fitparams}. $A_Q\{2\}$ (dotted curve) is then a {\em prediction} of $v_2$ measurements derived from the 1D projection onto azimuth of all 2D angular correlation structure, corresponding to measured $v_2\{2\} \approx v_2\{EP\}$~\cite{newflow,2004}. 

Figure~\ref{nonflow} (right panel) shows $v_2\{{\rm method}\}$ trends obtained from the corresponding $A_Q\{{\rm method}\}$ curves in the left panel via Eq.~(\ref{method}). 
Also included are $v_2\{2\}$ data (open squares) from Ref.~\cite{2004}. Good agreement between data and prediction (dotted curve) is evident. Thus, from 2D nonjet quadrupole and minijet measurements we accurately predict $v_2\{EP\} \approx v_2\{2\}$ published data. The prediction does not include small contributions to $v_2\{2\}$ from HBT and electron pairs (more significant for peripheral \aa collisions) which are excluded from $A_{2D}$ by the 2D model-fit procedure~\cite{daugherity}.  For statistically well-defined $v_2$ methods (e.g., $v_2\{2\} \approx v_2\{EP\}$) the jet contribution can be estimated accurately and ``flow'' can be distinguish from ``nonflow.''
This exercise for $m=2$ is generalized to higher multipoles in following sections. 

\subsection{1D ZYAM subtraction: $\bf m = 2$}

Dihadron angular correlations are projections onto 1D azimuth of 2D angular autocorrelations. Dihadron correlation analysis includes ``trigger-associated'' $p_t$ cuts intended to enhance jet structure relative to background. A nonjet combinatoric background is estimated by the {\em zero yield at minimum} or ZYAM procedure and subtracted from the sibling-pair density distribution to obtain an estimate of jet-related correlation structure. The background estimate relies on published $v_2(p_t)$ data~\cite{2004,dihadron}. 

Two problems arise from ZYAM subtraction: (a) the ZYAM offset estimate is not valid for overlapping peaks (as encountered in more-central \aa collisions) leading to large errors in the apparent zero offset and inferred peak shapes and (b) published $v_2(p_t)\{\text{method}\}$ data may include substantial contributions from the SS 2D jet peak (nonflow). The result is underestimation of jet fragment yields and distortion of inferred jet correlations~\cite{tzyam}.

 \begin{figure}[h]
   \includegraphics[width=3.3in,height=1.6in]{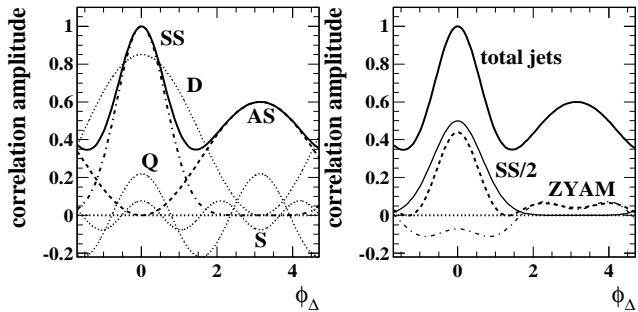}
\caption{\label{zyam1}
Left: Simulated dihadron correlation data (bold solid curve) for central $b = 0$ \auau collisions relative to true baseline, with SS 1D Gaussian (dash-dotted curve) and AS dipole (dashed curve). Dotted curves labeled D, Q and S are respectively the dipole, quadrupole and sextupole Fourier components of the SS 1D Gaussian.
Right: Original correlation data (bold solid curve, total jets) and bold dashed curve illustrating the result of ZYAM subtraction with biased background $v_2$ including 50\% of the SS 1D Gaussian (jet-related) quadrupole component. The dash-dotted curve illustrates the origin of the AS structure (see text).
 } 
 \end{figure}

Figure~\ref{zyam1} (left panel) shows simulated data from 200 GeV central ($b = 0$) \auau collisions (from Sec.~\ref{angcorrdata}) projected onto azimuth (bold solid curve). For central \auau collisions the nonjet quadrupole is consistent with zero~\cite{davidhq}. Correlation structure consists entirely of the SS 1D Gaussian (dash-dotted curve) and AS dipole (dashed curve). The zero offset is well-defined based on model fits to 2D histograms. Also shown are SS 1D peak (jet-related) multipoles D, Q,  S (dipole, quadrupole, sextupole). In 2D angular correlations the SS multipoles with their large curvatures on $\eta_\Delta$ are distinct from the AS dipole and nonjet quadrupole which have negligible curvatures within $|\eta| < 1$ (see App.~\ref{2dmult}).

Fig.~\ref{zyam1} (right panel) shows the result of ZYAM background subtraction (dashed curve).  Half the SS 1D peak quadrupole amplitude $v_2\{SS\}$ is typically included in the $v_2$ used for ZYAM subtraction  (nonflow bias)~\cite{tzyam}. The SS 1D Gaussian is then effectively divided into a half-amplitude Gaussian and ``liberated'' SS dipole and sextupole terms. The original data (total jets, bold solid curve) and a half-amplitude SS 1D peak (SS/2, lower solid curve) are shown for comparison. What remains after subtraction (dashed curve) is the sum of the half-amplitude SS Gaussian (SS/2), a half-amplitude sextupole (S/2) and a small-amplitude (negative) net dipole (from near cancelation of SS/2 + AS dipoles). The small net (negative) dipole is reinterpreted as ``bulk momentum conservation,'' and the S/2 SS sextupole produces relatively large distortions in the surviving AS structure interpreted as evidence for ``Mach cones''~\cite{mach}. The dash-dotted curve represents the sum S/2 + net dipole, demonstrating the origin of ZYAM-induced distortion of AS structure. Such distorted and suppressed jet manifestations are used to support claims for formation of an opaque bulk medium with anomalous properties in RHIC collisions. The distortions arising from $v_2$ oversubtraction in ZYAM analysis are worsened by newly-introduced analysis methods described in the following two sections.

\section{Triangular flow} \label{triangle}

In a follow-up to ZYAM subtraction the SS sextupole component ``released'' from the SS 1D Gaussian by $v_2$ oversubtraction is redefined as the FS manifestation of a ``triangularity'' component of IS geometry coupled to radial flow~\cite{gunther}.
The $m=3$ Fourier component of the SS 2D peak is then identified as ``triangular flow.'' The strategy eliminates remaining manifestations of the SS 2D jet peak as such {\em in the 1D projection}, in part by suppressing the critical \deta curvatures which distinguish SS jet structure from nonjet structure. 
In this and the following section we summarize recent arguments favoring ``higher harmonic flows'' $V_m$ in the final state. In Sec.~\ref{predict} we respond with $V_m$ predictions based on a two-component minijet-quadrupole model of FS correlation structure.

\subsection{Example: 1D Fourier analysis}

\begin{figure}[h]
  \includegraphics[width=3.3in,height=3.3in]{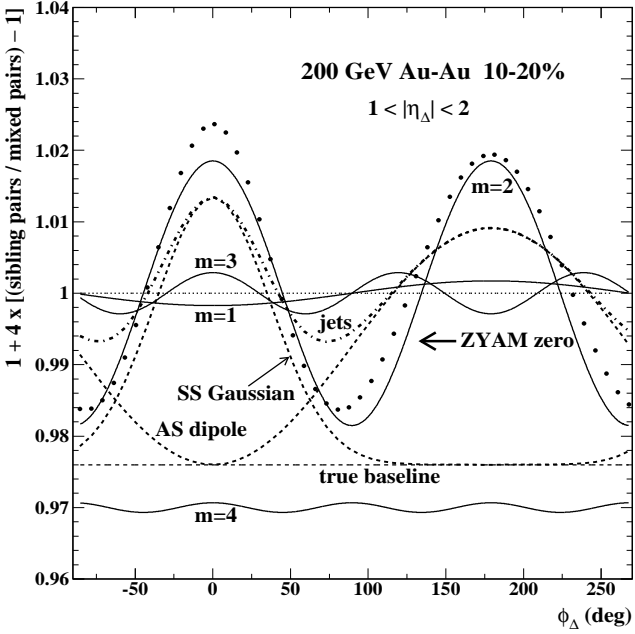}
\caption{\label{gunther}
Fourier analysis of azimuth correlations for 10-20\% central 200 GeV \auau collisions. Simulated data based on Ref.~\cite{daugherity} are shown as points projected from a 2D histogram. Jet structure inferred from 2D fits is shown as dashed and dash-dotted curves. Solid curves represent Fourier components similar to those derived in Ref.~\cite{gunther} but inferred in this case from fitted jet structure and nonjet quadrupole (see text). The curve labeled $m=4$ can be compared with the fit residuals in Ref.~\cite{gunther}.
 } 
 \end{figure}

Figure~\ref{gunther} is similar to the right panel of Fig.~1 from Ref.~\cite{gunther} which includes an analysis of STAR dihadron data. The points in Fig.~\ref{gunther} are simulated 10-20\% central 200 GeV \auau data derived from a STAR data analysis~\cite{daugherity}. The corresponding 2D histogram is similar to that shown in Fig.~\ref{2dcorr} (c) with a somewhat different centrality (10-20\% instead of 20-30\%). The 10-20\% parameters from Eq.~(\ref{estructfit}) are $A_{2D} = 0.77$, $2A_Q = 0.18$ and $A_D = 0.58$. The amplitude of the SS 1D Gaussian projected from 2D is  $A_{1D} = 0.85 \times 0.77 = 0.65$. Projection factor $G = 0.85$ (with $\eta$ exclusion cut, see App.~\ref{etacuts}) corresponds approximately to the restricted (``long range'') $\eta$ interval employed in Ref.~\cite{gunther} ($\eta$ cut noted in Fig.~\ref{gunther}).

The data points shown in Ref.~\cite{gunther} represent pair ratio $\langle r \rangle =  \rho / \rho_{ref}$ and are plotted on one quadrant of $(\eta_\Delta,\phi_\Delta)$. Thus, conversion of parameters $A_{1D}$, $A_Q$ and $A_D$ is required. The conversion factor from Eq.~(\ref{estructfit}) to $\Delta \rho / \rho_{ref} = \rho / \rho_{ref} - 1$ is $1 / \rho_0(b)$, with $\rho_0(\text{10-20\%}) \approx 70$. An additional factor 4 is required to convert from a single quadrant to the full angular acceptance. For the STAR data used in Ref.~\cite{gunther} a $p_t > 0.8$ GeV/c cut was applied, which results in a somewhat narrower (on azimuth) SS 2D peak with reduced amplitude.

Figure~\ref{gunther} also shows model elements for the SS 1D Gaussian and AS dipole (dashed curves) and their sum (dash-dotted curve labeled ``jets'') with the true baseline established by the 2D fit to data. Fourier components of the 1D data were derived in Ref.~\cite{gunther} as follows. A four-term Fourier series ($m  =0$-3) was fitted to the data to obtain amplitudes for dipole, quadrupole and sextupole terms. The Fourier fit was subtracted from the data to obtain residuals dominated by an $m=4$ octupole component.

Results from such a procedure can be anticipated from measured properties of the SS 2D Gaussian, AS dipole and nonjet quadrupole as follows. From Fourier coefficients $F_m(\sigma_{\phi_\Delta})$, with $\sigma_{\phi_\Delta} = 0.65$ for the SS 1D peak as determined in Sec.~\ref{periodpeak}, we obtain total quadrupole amplitude $4(2A_Q + F_2 A_{1D})/70 = 0.019$  (``flow'' + ``nonflow'' or nonjet + SS quadrupoles respectively), total dipole amplitude $4(2F_1 A_{1D} - A_D)/70 = -0.0017$ (SS + AS dipoles respectively) and SS sextupole $4F_3 A_{1D}/70 = 0.0029$. Residuals should be dominated by SS octupole amplitude $4 F_4 A_{1D}/70 = 0.0007$. The corresponding solid curves in Fig.~\ref{gunther} are consistent with the results presented in Fig.~1 of Ref.~\cite{gunther}. 

Reference~\cite{gunther} interprets results from the Fourier analysis (solid curves) to indicate that there is a large ``elliptic flow'' component ($m = 2$), a significant ``triangular flow'' component ($m  =3$) and a small negative dipole ($m = 1$) representing ``global momentum conservation.''  The description refers to ``long-range'' correlations, but examination of Fig.~\ref{2dcorr} (c) reveals that the difference between ``long-range'' ($|\eta_\Delta| > 1$) and ``short-range'' ($|\eta_\Delta| < 1$) projections is minor. Short-range correlations may reveal somewhat larger ``triangular flow.'' There is no acknowledgment of possible jet structure in minimum-bias angular correlations. 

In analysis based on 1D projections the large curvature on $\eta_\Delta$ of the SS 2D peak is suppressed. But that curvature distinguishes the SS 2D peak from the AS dipole and nonjet quadrupole, both approximately uniform on $\eta_\Delta$ within the relevant $\eta$ acceptance. Thus, the sums $2A_Q + F_2 A_{1D}$ and $2F_1 A_{1D} - A_D$ combine numbers belonging to different categories of 2D structure with misleading results. 
The Fourier components of the projected SS 2D peak cannot be considered individually (e.g., ``triangular flow'') because they {\em must} sum to zero in the AS azimuth region to describe measured 2D data. This approach marks a reversal of conventional attempts to discriminate ``elliptic flow'' (nonjet quadruple) from ``nonflow'' (jet-related quadrupole), for example with four-particle $v_2\{4\}$ or Lee-Yang zeroes methods~\cite{2004} (see App.~\ref{nonflowapp}).  

\subsection{Triangular flow centrality systematics}

\begin{figure}[h]
   \includegraphics[width=1.65in,height=1.65in]{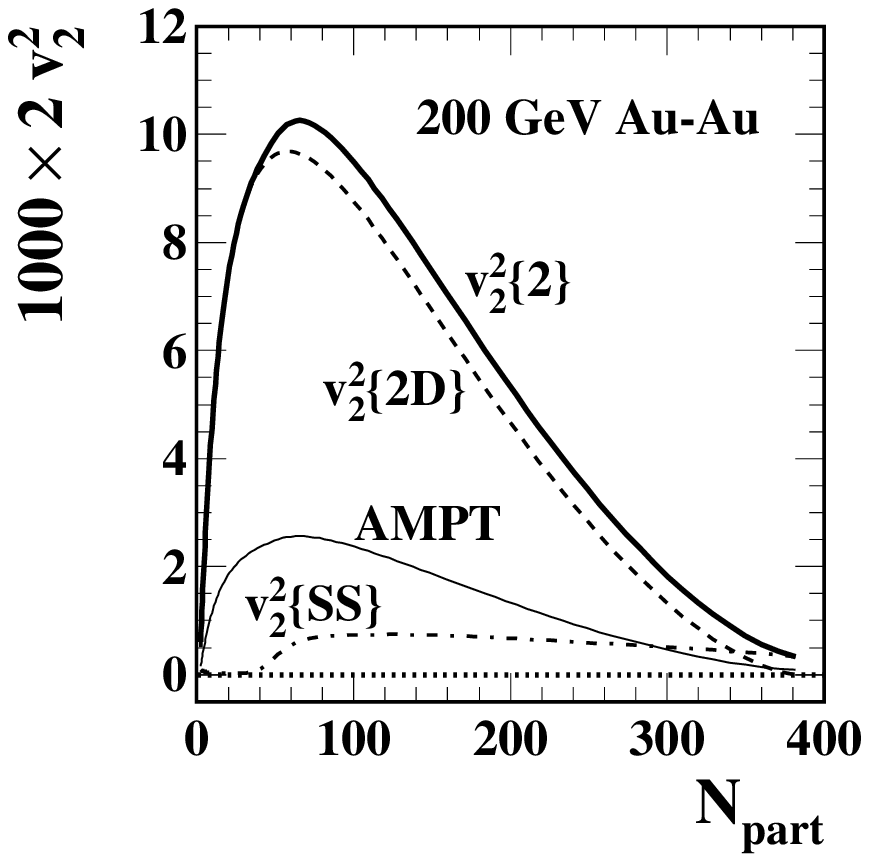}
   \includegraphics[width=1.65in,height=1.65in]{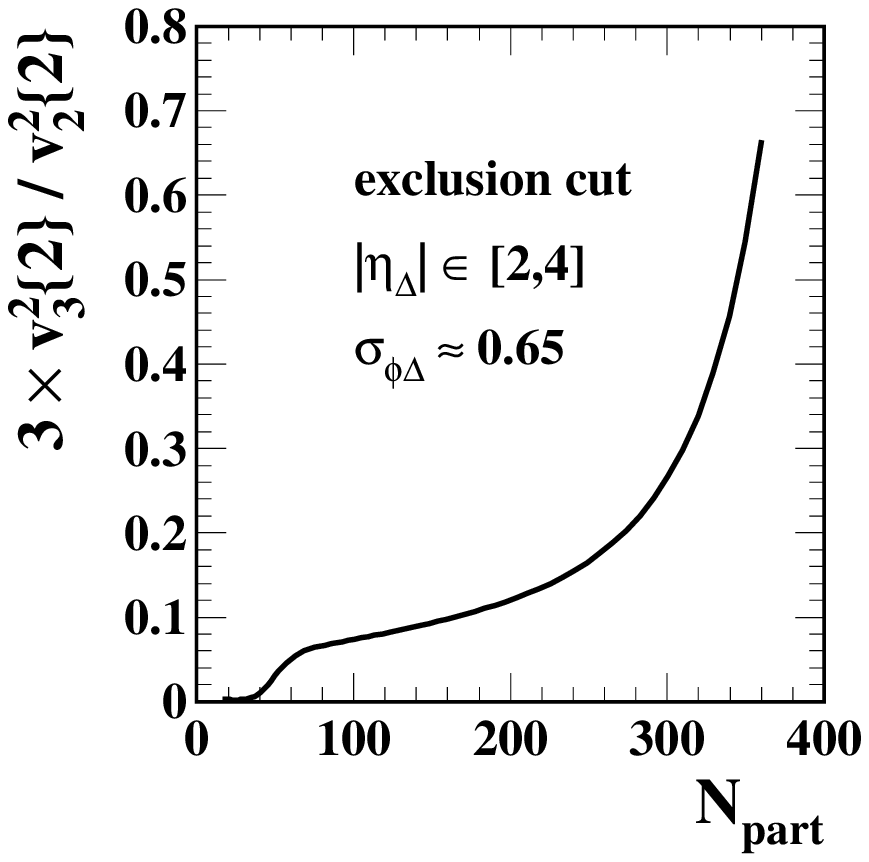}
\caption{\label{triang1}
Left: Comparison of three $v_2$ methods obtained from the parametrizations in Sec.~\ref{angcorrdata} with $\eta$ exclusion cut $|\eta_\Delta| \in [2,4]$ and plotted on $N_{part}$ as in Fig.~5 (left panel) of Ref.~\cite{gunther}.
Right: Ratio of sextupole to quadruple amplitudes vs $N_{part}$ comparable to  Fig.~8 (left panel) of Ref.~\cite{gunther}
 } 
 \end{figure}

Figure~\ref{triang1} (left panel) shows $v_2^2\{\text{method}\}$ plotted in the format of Ref.~\cite{gunther}. The curves are based on STAR minimum-bias angular correlations presented in Sec.~\ref{angcorrdata}. The SS 2D peak $\eta$ width plays the dominant role in systematic variations. 
This plot is a variant of Fig.~\ref{nonflow} (right panel), but the $\eta$ exclusion cut applied in this case ($|\eta_\Delta| \in [2,4]$) insures much stronger ``extinction'' of the SS 2D jet peak contribution in more-peripheral \aa collisions (below the sharp transition at $\nu \approx 3$ or $N_{part} < 50$) than the cut  $|\eta_\Delta| \in [1,2]$ applied in Fig.~\ref{gunther} and Sec.~\ref{vmetacuts}.

Figure~\ref{triang1} (right panel) shows ratio $v_3^2\{SS\} / v_2^2\{2\}$ which includes contributions from the SS peak in both numerator and denominator but contributions from nonjet quadrupole $v_2^2\{2D\}$ in the denominator only. The ratio also falls to zero below $N_{part} = 50$, because the only contribution to numerator $v_3^2\{SS\}$ is from the SS 2D peak.  With increasing centrality $v_2^2\{2D\}$ in the denominator falls to zero and the ratio must increase to the limit $3 F_3 / F_2 \approx 1$ consistent with 2D correlation data.

 \begin{figure}[h]
   \includegraphics[width=1.65in,height=1.65in]{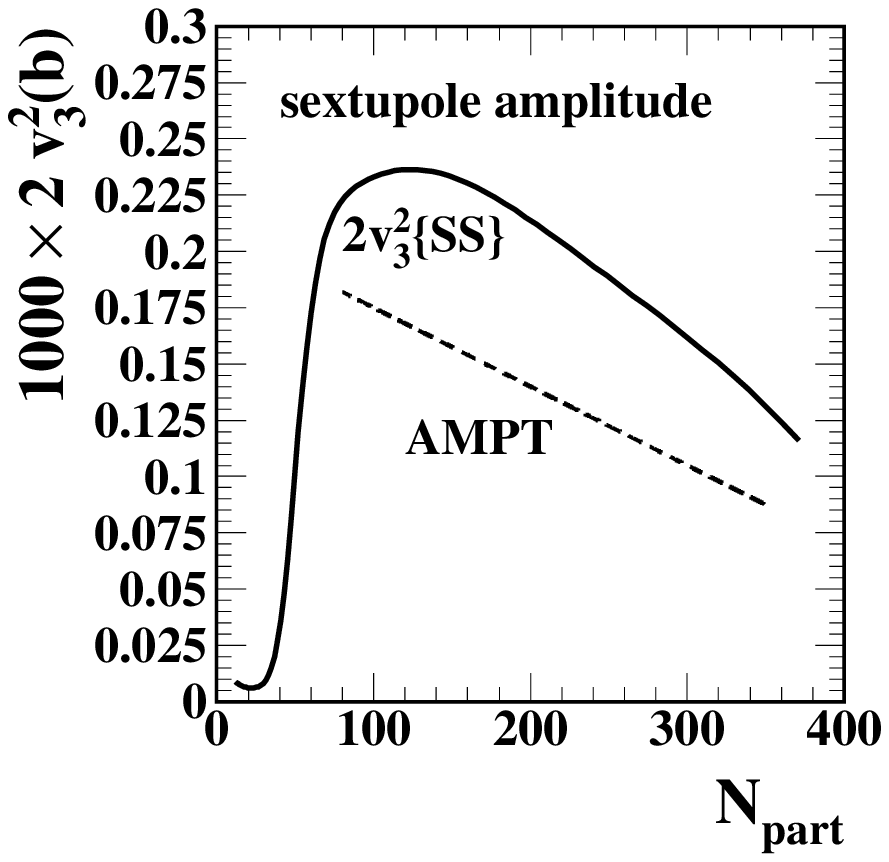}
   \includegraphics[width=1.65in,height=1.65in]{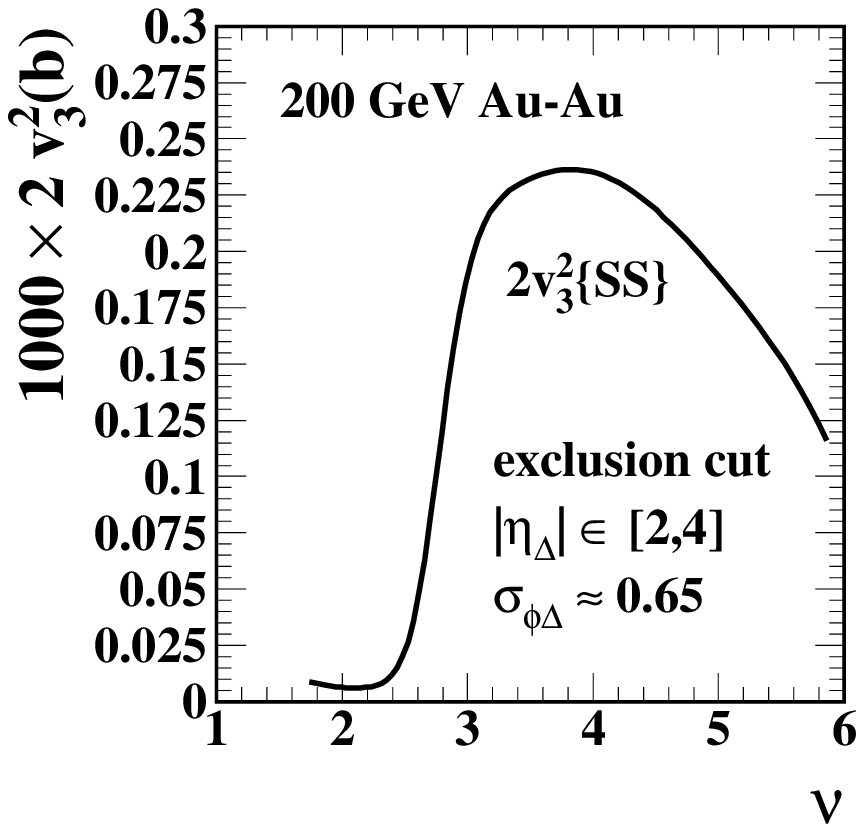}
\caption{\label{triang2}
Left: Sextupole amplitude obtained from the parametrizations in Sec.~\ref{angcorrdata} with $\eta$ exclusion cut $|\eta_\Delta| \in [2,4]$ and plotted on $N_{part}$ as in Fig.~5 (right panel) of Ref.~\cite{gunther}. 
Right: The same result plotted on participant path length $\nu$, showing the correspondence to the sharp transition in SS 2D peak properties in Fig.~\ref{fitparams}, especially the $\eta$ elongation.
 } 
 \end{figure}

Figure~\ref{triang2} shows $v_3^2\{SS\}$ for two plotting formats comparable to Fig.~5 (right panel) of Ref.~\cite{gunther}.  Note that $v_m^2\{2\} = v_m^2\{SS\}$ for $m = 3$. The sextupole component becomes nonzero only above the sharp transition in SS 2D peak properties where the SS peak elongates on $\eta$ into the ``ridge-like'' $\eta_\Delta$ acceptance.  Per-pair measure $v_3^2\{SS\}$ decreases strongly above the sharp transition at $\nu \approx 3$. In contrast, per-particle sextupole amplitude $A_S\{SS\}$ would continue to increase monotonically with centrality, reflecting increased jet production.  

The results from this demonstration based on previously-measured minijet 2D angular correlations are generally duplicated by results from Ref.~\cite{gunther}. Reported ``triangular flow'' measurements are consistent with the properties of a monolithic SS 2D jet peak.  Further details are presented in Sec.~\ref{predict}. The effects of $\eta$ exclusion cuts are considered in detail in Sec.~\ref{vmetacuts}.

\section{Long-range Correlations} \label{luzumsec}

In a followup to triangular flow Ref.~\cite{luzum} disputes  a claim in Ref.~\cite{starzyamnew} that dependence on the reaction-plane angle of ZYAM-subtracted long-range (on $\eta$) azimuth correlations signals nonflow. Instead, all such ``ridge-like'' correlations should be attributed to collective flows, and ZYAM-subtracted data are said to be consistent with that hypothesis:
``...the measured dihadron correlation at large $\Delta \eta$ consists almost entirely of the lowest few Fourier components, each of which can be quantitatively understood as coming from collective flow (plus global momentum conservation)...''~\cite{luzum}.
Long-range azimuth correlations are said to originate at early times and therefore must represent the ``collective behavior of the system'' (and global momentum conservation). Here we present evidence contradicting claims in Ref.~\cite{starzyamnew} and Ref.~\cite{luzum}.

\subsection{Fourier series analysis of long-range structure} \label{luzfour}

A Fourier series analysis of unsubtracted ``ridge-like'' trigger-associated 1D azimuth correlations (with trigger related to the event plane) is shown in Fig.~1 of Ref.~\cite{luzum}.  $\eta$ exclusion cuts $0.7 < |\eta_\Delta| < 2$ reject a short-range or jet-like part of the pair acceptance on $\eta_\Delta$ in favor of a long-range or ridge-like part of the acceptance. The analysis reports four significant Fourier coefficients $V_m\{2\}$ ($m = 1$-4), where $V_m = v_m^a v_m^t$ is nominally a product of trigger and associated $v_2(p_t)$ values. The $V_2\{2\}$ quadrupole term dominates mid-central \auau collisions. 
Absence of higher harmonics ($m > 4$) is considered remarkable
but should be expected given the widths of peaked structures observed by previous analyses~\cite{axialci,daugherity,tzyam}, as discussed in Sec.~\ref{periodpeak}.  

Inferred quadrupole and octupole terms $V_2\{2\}$ and $V_4\{2\}$ are attributed solely to elliptic flow. The amplitudes are said to vary relative to event-plane angle as $\cos(2\phi_s)$ and $\cos(4\phi_s)$ as expected for flows, with $\phi_s = |\phi_t - \psi_{EP}|$ (trigger and event-plane angles). Fitted $V_2\{2\}$ values are said to exceed ``measured'' $V_2\{ZA\}$ data (estimated $V_2$~\cite{starzyamnew}) by a significant amount (attributed to nonflow). But since the nonflow {\em appears to have the same dependence} on $p_t$ and $\phi_s$, $V_2\{ZA\}$ must underestimate the true flow, and $m=2,4$ ``nonflow'' must be flow.

Conjectured flow terms $V_1\{2\}$ and $V_3\{2\}$ are not expected to correlate with the reaction plane, and the Fourier analysis in Ref.~\cite{luzum} is interpreted to confirm that expectation. The $m=3$ sextupole term increases monotonically with $p_t$. It is therefore concluded that the $m=1,3$ terms are also consistent with flows. Thus, all long-range or ``ridge-like'' structure must be flows.

However, the ``nonflow'' components of $V_2\{2\}$ and $V_4\{2\}$ (excesses over corresponding ZA estimates) do not actually follow the sinusoid trends expected for flows (see following subsections). $V_3$ is said be independent of $\phi_s$, implying a flow interpretation, but the $V_3$ data from four of five $p_t$ intervals show significant increases, and real ``nonflow'' (jets) may or may not have a significant dependence on $\phi_s$. There is insufficient sensitivity to details of $p_t$ dependence that might distinguish ``flows'' from jet structure. Thus, the results in Fig.~1 of Ref.~\cite{luzum}, examined {\em sufficiently differentially}, are either inconclusive or {\em actually confirm the presence of nonflow} (jets) in the dihadron data. In the rest of this section we examine the case for $m=2$ in detail.

\subsection{Event-plane-related ZYAM subtraction}

The ZYAM discussion in Ref.~\cite{luzum} is based on an analysis of dihadron correlations which included free fits with a model function applied to dihadron data as a check on the ZYAM subtraction~\cite{starzyamnew}. The fit model
\bea \label{estruct}
\frac{dN_{pair}}{d\phi_\Delta} &=& B\left\{  1 + 2V_2\cos(2\phi_\Delta) + 2V_4 \cos(4\phi_\Delta)   \right\} \\ \nonumber
&+& A_{1D}  \, e^{- \frac{1}{2}  \left( {\phi_{\Delta}}/{ \sigma_{\phi_{\Delta}}} \right)^2  } + A_D \cos(\phi_\Delta - \pi)
\eea
is similar to Eq.~(\ref{estructfit}) of the present analysis. 
Based on the fit results Ref.~\cite{starzyamnew} concluded that free model fits must be incorrect for two reasons:
(a) ``In order to eliminate the away-side double-peak...flow [fitted $V_2$, $V_4$] that is much larger than {\em experimentally measured} [$V_m\{ZA\}$ defined below] is required....'' and (b) ``...deviations of the {\em fitted} flow modulations from the {\em measured} ones vary slice to slice [$\phi_s$ bins], which should not be the case if the {\em measured} flow parameters...were simply in error'' [emphasis added].  Thus, ``...the fit model [Eq.~(\ref{estruct})]...cannot be the correct functional form to describe the dihadron correlation signal....''
But that conclusion doesn't follow from the fitting exercise, which actually demonstrates that the unsubtracted dihadron data are well described by Eq.~(\ref{estruct}). 

The values of $V_2\{ZA\}$ used to define the background for event-plane-related ZYAM subtraction are derived from Eq.~4 of Ref.~\cite{starzyamnew}, which assumes prior knowledge of the product $v_2^{(a)}(p_t)v_2^{(t)}(p_t)$ for trigger and associated \pt bins and the event-plane resolution from separate numerical analysis. There is reason to question such data based on independent 2D correlation analysis~\cite{davidhq2}.

 \begin{figure}[h]
  \includegraphics[width=3.3in,height=3.3in]{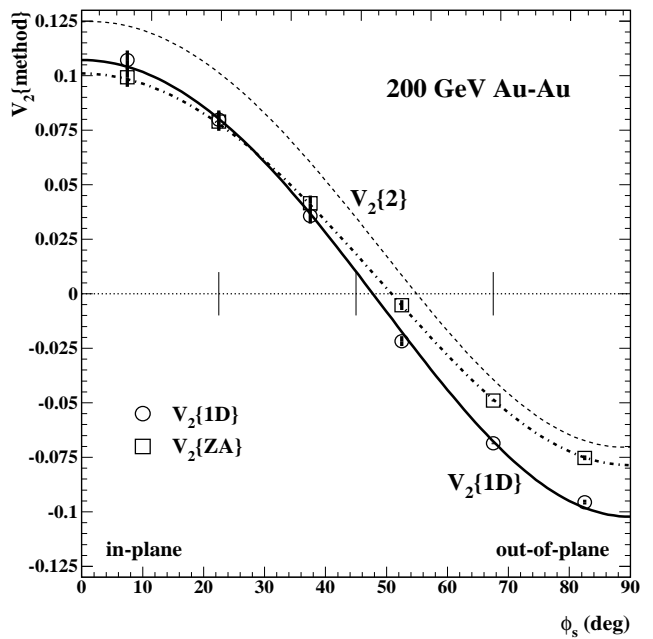}
\caption{\label{badzyam}
Measured quadrupole amplitudes from two methods. $V_2\{1D\}$ data (open circles) are derived from free model fits to 1D azimuth correlations, and $V_2\{ZA\}$ data (open squares) are estimated from published $v_2$ data~\cite{starzyamnew}. Error bars are shown within the open symbols. The $V_2\{2\}$ trend (dashed curve) is based on an estimate of the SS 2D peak contribution $V_2\{SS\}$ summed in quadrature with $V_2\{1D\}$.
Thin vertical lines mark zeros of $\cos(2\phi_s)$ and $\cos(4\phi_s)$. The dash-dotted and solid curves are both obtained from Eq.~(4) of Ref.~\cite{starzyamnew} (see text).
 } 
 \end{figure}

Figure~\ref{badzyam} is similar to Fig.~1 (left column, second panel) of Ref.~\cite{luzum}. The figure shows ``measured'' $V_2\{ZA\}$ values (ZYAM analysis $\rightarrow$ ZA) reported in Table II of Ref.~\cite{starzyamnew} (open squares, note error bars within) based on parameters in Table I.  The dash-dotted curve is Eq.~4 of Ref.~\cite{starzyamnew} with the same parameters from Table I inserted, confirming consistency.
The ``fitted'' $V_2\{1D\}$ values inferred from 1D free fits (reconstructed here from correction factors provided in Ref.~\cite{starzyamnew}) are shown as open circles.  The free-fit results {\em also constitute measurements}, but derived directly from fits to the dihadron data in question rather than indirectly from separate analysis of other particle data.

Variation of correction {\em factors} ``from slice to slice'' (a) may be irregular due to the rapidly changing magnitude (and sign) of $V_2$, but the vertical {\em offset} from one data trend to another varies slowly and smoothly.
``Much larger'' $V_2$ values (b) are not observed.  {Absolute} changes in inferred quadrupole amplitudes are actually modest, within typical systematic uncertainties for $v_2\{EP\}$ data (see below). The solid curve is Eq.~(4) of Ref.~\cite{starzyamnew} with the event-plane resolution increased by 10\% and the trigger $v_2$ reduced from 0.16 to 0.1, consistent with Ref.~\cite{davidhq2}.

It is important to note that the small difference between the solid curve and the dash-dotted curve in Fig.~\ref{badzyam} accounts entirely for the difference between jet structure {\em undistorted and increasing in amplitude} from in-plane to out-of-plane and jet structure {\em increasingly distorted and suppressed} from in-plane to out-of-plane from Ref.~\cite{starzyamnew}. 

\subsection{Argument for reinterpretation of ZYAM results}

ZYAM subtraction is intended to remove ``elliptic flow'' and an uncorrelated combinatoric background from dihadron azimuth correlations. Reference~\cite{starzyamnew} argues that since ZYAM-subtracted ``ridge-like'' correlations depend on reaction-plane angle  $\phi_s$ (e.g., increasing suppression of jet-like structure) the surviving structure must represent ``nonflow,'' since recently-proposed higher odd {\em flow} harmonics (e.g., $v_3$) should not relate to the reaction plane. 

Based on its own inferred $\phi_s$ and $p_t$ trends Ref.~\cite{luzum} responds that 
additional flow structure must remain after ZYAM subtraction, the amount depending on $\phi_s$. 
ZYAM $v_m$ ``should come from an independent measurement that does not contain a contribution from non-flow correlations.... Such a measurement does not exist.''
Because the $v_m$ are not known ``...the result [of ZYAM subtraction] will have significant contributions from flow.'' In other words, unknown systematic $v_m$ biases must be such that measured $v_m$ {\em always underestimate} ``real'' flows by overestimating nonflow. The upper limit estimated by $V_2\{ZA\}$ is therefore too low. 
One should impose a Fourier analysis on all long-range or ``ridge-like'' correlations to obtain $V_m\{2\}$ representing ``real'' flows. With that definition no nonflow (jets) could survive ``background'' subtraction---all jet structure must be redefined as flows.

\subsection{Estimating the nonflow contribution}

In Ref.~\cite{luzum} we identify the term ``estimated flow contribution'' with $V_2\{ZA\}$ and ``extracted Fourier component'' with fitted $V_2\{2\} \approx V_2\{EP\}$, which is said to measure ``real'' flow. The difference $V_2\{2\} - V_2\{ZA\}$ is defined as ``nonflow,'' which is claimed to be overestimated based on inferred $\phi_s$ and $p_t$ trends. 
Reference~\cite{luzum} concludes that nonflow should be reduced to zero and $V_2\{2\}$ adopted as ``real'' flow to be used in ZYAM subtraction, along with other higher harmonics similarly determined. The concept is illustrated in Fig.~2 of Ref.~\cite{luzum}, with contrasting subtraction results from $V_2\{ZA\}$ (left panels) and from $V_2\{2\}$ (right panels).

Based on study of 2D correlations and their 1D projections we find that $V_2\{2\} = V_2\{SS\} + V_2\{1D\}$~\cite{davidhq,davidhq2,gluequad}. And $V_2\{SS\}$ is quantitatively predicted by the SS 2D peak with its large curvature on $\eta_\Delta$, whereas $V_2\{1D\}\approx V_2\{2D\}$ represent an azimuth sinusoid with no curvature on $\eta_\Delta$. 
We can therefore predict $V_2\{2\}$ from Ref.~\cite{luzum}  based on the free-fit results and the quadrupole component of the measured SS 2D jet peak (nonflow) as follows.  The SS peaks derived from free fits and shown in Fig.~4 of Ref.~\cite{starzyamnew} (dashed curves) increase in amplitude with $\phi_s$ smoothly over the interval 0.7-1.2. The background constant for the 1-2 GeV/c $p_t$ cut interval is $B \approx 3.8$. Given Fourier coefficient $F_2 \approx 0.2$ we obtain for the SS peak quadrupole $2V_2\{SS\} = 0.2 (\text{0.7-1.2}) / 3.8$ or $V_2\{SS\} \approx$ 0.02-0.03. Combined with $V_2\{1D\}$ from the free fit (solid curve) in Fig.~\ref{badzyam} we obtain $V_2\{2\}$ as the dashed curve, which is consistent with Fig.~1 of Ref.~\cite{luzum}. Positive-definite nonflow $V_2\{SS\}$ (jet-related quadrupole) increases monotonically with $\phi_s$, a trend inconsistent with flow expectations.

Given the structure of 2D angular correlations, nonflow relevant to ZYAM is more properly defined as  $V_2\{ZA\} - V_2\{1D\} $, the excess of ZYAM $v_2$ over the nonjet (fitted) quadrupole value. The nonflow upper limit is then  $V_2\{2\} - V_2\{1D\}$. Subtracting $V_2\{2\}$ would achieve complete removal of the SS peak from projected 1D dihadron correlations. But the SS peak remains the most prominent structure in 2D angular correlations and cannot be removed by such 1D subtraction procedures. 
Reduction of nonflow bias to zero in the $V_2$ estimate is actually achieved by invoking $V_2\{1D\}$ from the free fits.

\section{Predicting $\bf v_m$ measurements} \label{predict}

Based on model fits to 2D histograms, including minijet and nonjet-quadrupole structure, we can predict any $v_m$ measurement for any $\eta$ exclusion cuts. In contrast, 1D projections onto $\eta_\Delta$ or $\phi_\Delta$ cannot predict 2D structure because the projections abandon essential information. Nongraphical numerical methods (e.g., conventional $v_m$ measurements) and dihadron correlations depend explicitly or implicitly on 1D projection followed by a single Fourier-series decomposition of the azimuth projection. 

We generalize the ``flow vs nonflow'' case for $m=2$ to higher $m$ by projecting all measured 2D structure to 1D and invoking a single Fourier series to predict recent $v_m\{2\}$ measurements. We introduce $\eta$ exclusion cuts to simulate attempts to distinguish (remove) jet structure (nonflow) from conjectured hydro phenomena. This treatment emphasizes 1D projections onto azimuth. Extension to 2D correlations is described in App.~\ref{2dmult}.

\subsection{$\bf \eta_\Delta$ dependence of inferred $\bf v_m$}

Measurements of the $\eta$ dependence of ``triangular flow'' and other higher multipoles can be simply predicted from the 2D correlation model of Sec.~\ref{angcorrdata}. Since all higher multipoles are derived solely from the SS 2D peak, its parameters determine all multipole systematics.

 \begin{figure}[h]
\includegraphics[width=1.65in,height=1.6in]{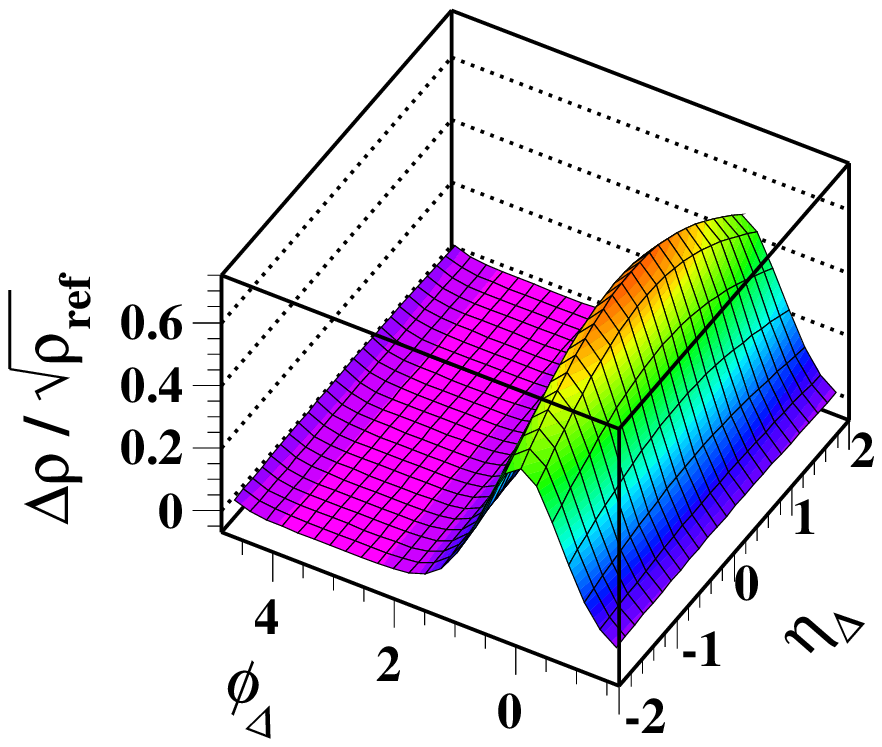}
  \includegraphics[width=1.65in,height=1.6in]{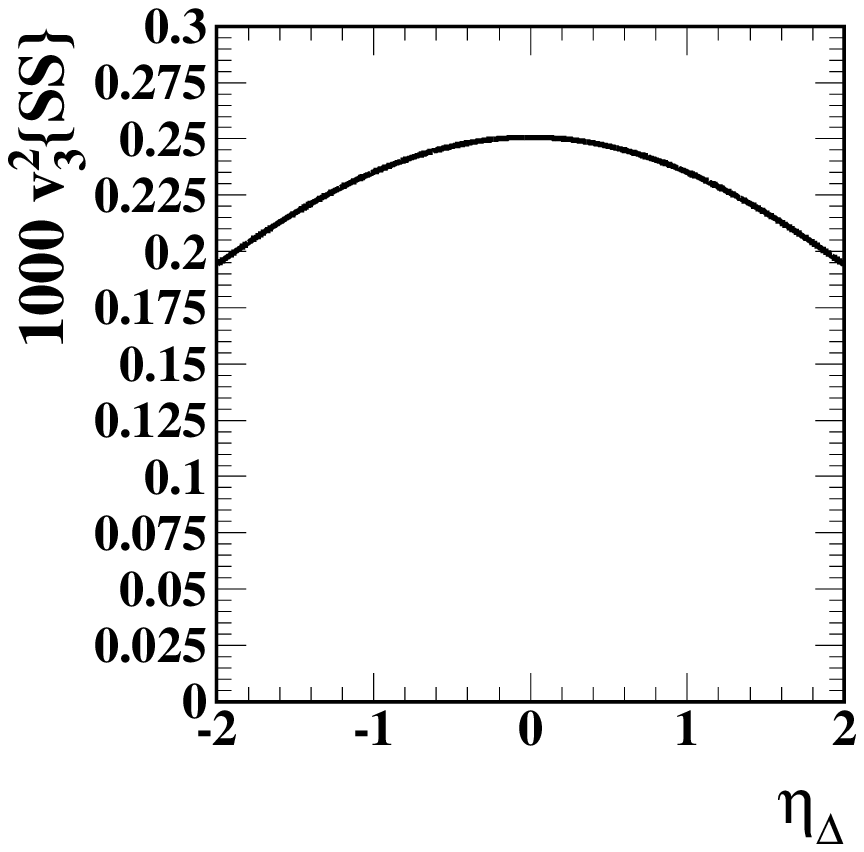}
\caption{\label{project} (Color online)
Left: Simulated angular-correlation data from 0-5\% central 200 GeV \auau collisions. An AS dipole component has been removed. The remainder is the SS 2D peak well-described by a 2D Gaussian.
Right: The sextupole component of the SS 2D Gaussian from the left panel determined as a function of $\eta_\Delta$, predicting what would result from such a ``triangular flow'' $v_3$ analysis applied to the same particle data.
 } 
 \end{figure}

Figure~\ref{project} (left panel) shows a correlation model of the SS 2D peak for 0-5\% central 200 GeV \auau collisions. It is the histogram in Fig.~\ref{2dcorr} (d) with the AS dipole removed.
Fig.~\ref{project} (right panel) shows the corresponding $v_3^2\{SS\}(\eta_\Delta)$ defined by
\bea
&& \hspace{-.3in} 2 \rho_0(b) v_3^2\{SS\}(\eta_\Delta)\hspace{-.0in} =\hspace{-.0in} F_3(\sigma_{\phi_\Delta}) A_{2D} \exp\left\{-\frac{\eta_\Delta^2}{2 \sigma^2_{\eta_\Delta}}\right\}.
\eea
Note that the mean value of that trend is consistent with the value of $v_3^2\{SS\}$ for most-central collisions from 1D projection onto azimuth in Fig.~\ref{multipoles} (right panel).

\subsection{$\bf v_m$ centrality trends in contiguous $\bf \eta$ acceptance}

The centrality dependence of any multipole integrated over a {\em contiguous $\eta$ acceptance} symmetric about the origin denoted by $\Delta \eta$ can be predicted from the 2D data model described in Sec.~\ref{angcorrdata}. The multipoles derived from the SS 2D peak are given by Eq.~(\ref{aqss}).

 \begin{figure}[h]
  \includegraphics[width=1.65in,height=1.65in]{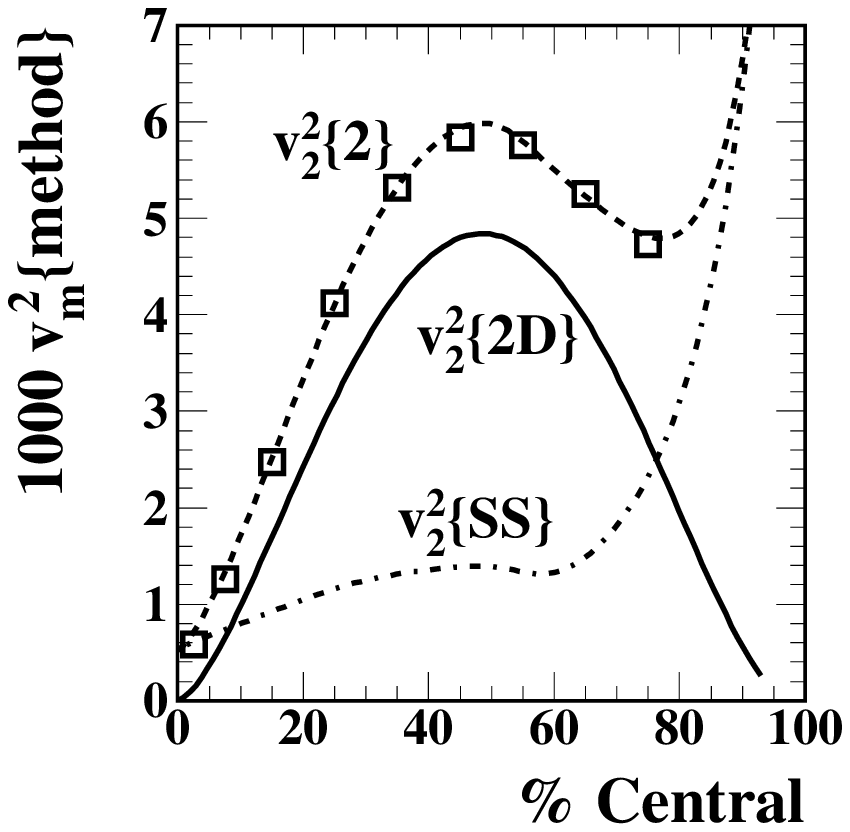}
  \includegraphics[width=1.65in,height=1.65in]{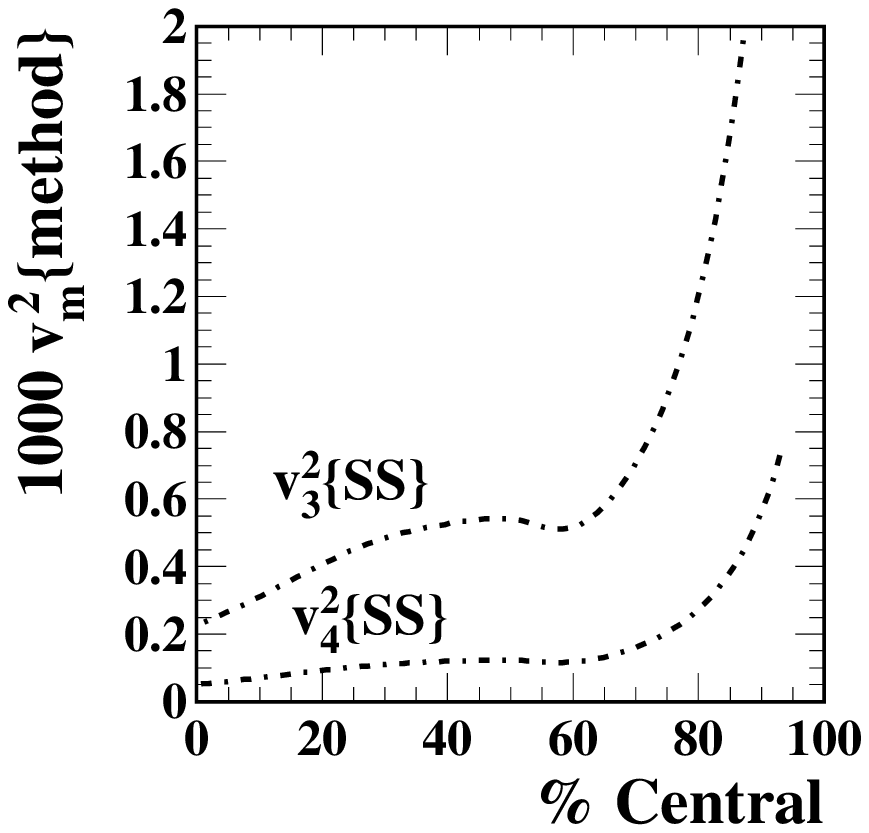}
\caption{\label{multipoles}
Left: Quadrupole components from 200 GeV \auau data vs centrality satisfying the relation $v_2^2\{2\} = v_2^2\{2D\}+v_2^2\{SS\}$~\cite{gluequad}, where $v_2^2\{SS\}$ is the jet-related quadrupole component of the SS 2D peak.
Right: Corresponding higher multipoles of the SS 2D peak, predicting the result of ``higher harmonic flow'' analysis applied to the same particle data.
 } 
 \end{figure}

Figure~\ref{multipoles} shows quadrupole, sextupole and octupole amplitude predictions in the form $v_m^2\{\text{method}\}(b)$ for 200 GeV \auau collisions, where SS (dash-dotted curves) indicates a multipole derived from the SS 2D peak, and 2D (solid curve) refers to the nonjet quadrupole inferred from 2D fits to data (Sec.~\ref{2dquad}). All SS multipoles are predicted to have the same $\eta_\Delta$ and centrality dependence, which would then confirm a common source (SS 2D peak). The common SS trends are very different from the unique nonjet quadrupole $v_2^2\{2D\}$ contribution. 
The good agreement between published data (open squares) and the predicted $v_2^2\{2\}$ trend (dashed curve) is notable, since the prediction combines two independent measurement programs (minijet systematics~\cite{daugherity} and nonjet quadrupole systematics~\cite{davidhq}) unrelated to the published data analysis~\cite{2004}.

\subsection{$\bf v_m$ centrality trends with $\bf \eta$ exclusion cuts} \label{vmetacuts}

It is argued that SS angular correlations can be separated by cuts on $\eta$ into a ``short-range'' or ``jet-like'' region at smaller $\eta_\Delta$ and a ``long-range'' or ``ridge-like'' region at larger $\eta_\Delta$~\cite{gunther}.  Structure in the ridge-like region is attributed to nonjet mechanisms, for instance initial-state geometry coupled with radial flow.
We derive an expression for the effect of $\eta$ exclusion cuts in App.~\ref{etacuts} assuming that particle pairs are formed from symmetric disjoint $\eta$ intervals $[-\eta_2,-\eta_1]$ and $[\eta_1,\eta_2]$ to obtain projection factor $G_m(\sigma_{\eta_\Delta};\eta_1,\eta_2)$. Multipole amplitudes are then defined in terms of 2D model elements by
\bea
&& \hspace{-.3in} 2 \rho_0(b) \, v_m^{\prime 2}\{SS\} \hspace{-.00in} = \hspace{-.00in} F_m(\sigma_{\phi_\Delta}) G_m(\sigma_{\eta_\Delta};\eta_1,\eta_2) A_{2D}(b),
\eea
the prime indicating an $\eta$ exclusion cut.

 \begin{figure}[h]
  \includegraphics[width=3.3in,height=3.3in]{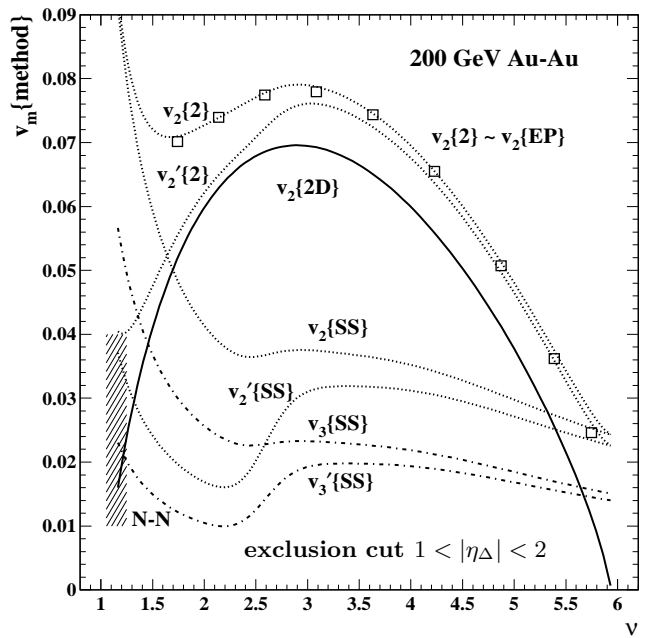}
\put(-150,30){\bf exclusion cut $1 < |\eta_\Delta| < 2$}
\caption{\label{etacuts1}
Illustration of ``nonflow'' reduction by $\eta$ cuts. The bold solid curve is the parametrized nonjet  quadrupole trend from Ref.~\cite{davidhq}. For each pair of remaining curves the upper curve corresponds to the full pair acceptance $|\eta_\Delta| < 2$ (as in Fig.~\ref{multipoles}) whereas the lower curve of each pair corresponds to a reduced pair acceptance $1< |\eta_\Delta| < 2$ intended to reduce nonflow. The $v_m\{SS\}$ trends are determined by the SS 2D peak properties presented in Fig.~\ref{fitparams}, with and without $\eta$ cuts.
} 
 \end{figure}

Figure~\ref{etacuts1} shows multipoles from the full $\eta$ acceptance, as in the previous subsection (upper curve for each line style), and from $\eta$ exclusion cuts with $\eta_1 = 0.5$ and $\eta_2 = 1$ (lower curve for  each line style) which exclude the ``short-range'' interval $ |\eta_\Delta| < 1$. The solid curve shows nonjet quadrupole $v_2\{2D\}$ which is independent of such cuts. The trends reflect smooth elongation of the SS 2D peak on $\eta_\Delta$ with increasing centrality, modulo a sharp transition in SS peak properties just below $\nu = 3$. 

The $p_t$-integral SS 2D peak shows no indication of being composite (e.g., distinct jet and ridge components on $\eta_\Delta$) for any \auau centrality. The SS peak is described in all cases by a single 2D Gaussian. The difference between $v_m$ and $v_m^\prime$ is entirely due to the changing SS peak $\eta_\Delta$ width relative to the special $\eta$ cut. More of the SS peak is  accepted by the $\eta$ exclusion cut in more-central collisions. The apparent extent of SS peak rejection (extinction) for more-peripheral collisions is minimized by plotting square root $v_m^\prime$ rather than $v_m^{\prime 2}$ which measures the actual correlated {\em pair} number. 
Aside from $\eta$ elongation all aspects of the SS 2D peak conform to pQCD jet expectations, including modification of fragmentation functions~\cite{fragevo}. 
Note that in Figs.~\ref{triang1} and \ref{triang2} the $\eta$ exclusion cut  is extended to $2 <|\eta_\Delta| <4 $ leading to more complete extinction of the SS peak contribution for more-peripheral \aa collisions. Response of the $V_m\{SS\}$ to $\eta$ exclusion cuts is a {\em very indirect} way to study the evolving $\eta$ structure of  the SS 2D peak.

\subsection{Comparison with recent LHC results}

\begin{figure}[h]
   \includegraphics[width=3.3in,height=3.3in]{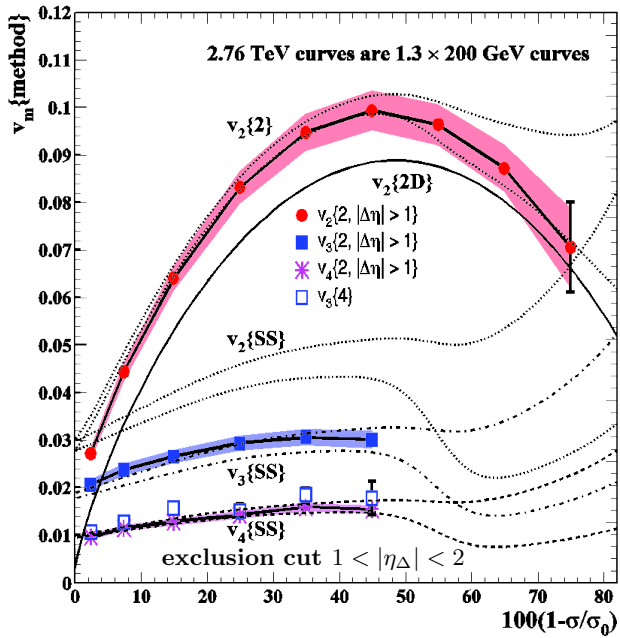}
\put(-180,26.5){\bf exclusion cut $1 < |\eta_\Delta| < 2$}
\caption{\label{alice} (Color online)
The data points are from Ref.~\cite{alice}. The various curves are obtained from Fig.~\ref{etacuts1} with $\sigma_{\phi_\Delta}$ reduced from 0.65 to 0.60 and with an overall multiplier 1.3. The relative magnitudes and centrality variations agree closely between data and curves, suggesting that the SS 2D peak at 2.76 TeV has similar properties to that at 200 GeV, as also noted in Ref.~\cite{ppcms}.
 } 
 \end{figure}

A test of $v_m$ predictions from the 200 GeV 2D data model is provided by comparison with a recent LHC analysis which infers ``higher harmonic flows'' from \pbpb data~\cite{alice}.
Figure~\ref{alice} shows $v_m\{2\}$ measurements from 2.76 TeV \pbpb collisions (points from Fig.~1 of Ref.~\cite{alice}) compared to the 200 GeV results summarized in Fig.~\ref{etacuts1} (curves). 
 The data from Ref.~\cite{alice} represent coefficients from a Fourier-series fit to the sum of all two-particle angular correlations projected onto 1D azimuth with $\eta$ exclusion cut $1 <|\eta_\Delta| <2$, denoted $v_m\{2,\text{$\eta$ cut}\}$.

The curves in Fig.~\ref{alice} are derived from Fig.~\ref{etacuts1} as follows. The SS peak azimuth width $\sigma_{\phi_\Delta}$ is reduced from 0.65 (200 GeV) to 0.60 (2.76 TeV). Azimuth width reduction is consistent with the energy trend from 62 to 200 GeV~\cite{daugherity}, increasing the $v_3$ and $v_4$ trends slightly relative to $v_2$. Given the SS peak width adjustment all curves in  Fig.~\ref{etacuts1} are multiplied by a common factor 1.3, the same factor attributed in Ref.~\cite{meanpt} to an increase in the inclusive spectrum mean $p_t$ at the higher energy. For each line style the upper curve corresponds to a contiguous $|\eta_\Delta| < 2$ acceptance while the lower curve corresponds to the nonflow exclusion cut  $1 <|\eta_\Delta| <2$.

Given those adjustments the predictions derived from 2D angular correlations at 200 GeV describe the \pbpb data at 2.76 TeV very well. The $v_2\{2\}$ trend follows the expectation for applied $\eta$ exclusion cuts, suggesting persistent presence of a sharp transition~\cite{daugherity} in the SS 2D peak $\eta_\Delta$ width at the higher energy. 
The lack of $v_3$ and $v_4$ data for more-peripheral centralities is unfortunate, because such data might provide a more direct test of SS 2D peak systematics at the higher energy. The LHC data are also consistent with the $v_2\{2D\}$ nonjet quadrupole trend (solid curve) which shows no sensitivity to $\eta$ exclusion cuts because the corresponding correlation structure is uniform on $\eta_\Delta$ within the $\eta$ acceptance.

 \section{Discussion}

This article addresses a prominent issue in the study of angular correlations at the RHIC: do jets or flows dominate the dynamics of high-energy heavy ion collisions? A trend has emerged recently to reinterpret jet angular correlations by projecting some part of the $\eta$ acceptance onto 1D azimuth and expressing the projection as a Fourier series. The series elements may then be interpreted as flows (or global momentum conservation). The FS multipoles are in turn attributed to conjectured IS geometry fluctuations coupled to radial flow. We are obliged to question the validity of isolating individual Fourier components of a single 2D peak structure, no matter what its interpretation. We should also question hydro interpretation of any multipole, given measured systematics of the nonjet quadrupole, AS dipole and SS 2D peak~\cite{davidhq}.

\subsection{Jet-related 2D angular correlation structure}

Correlations attributable to minimum-bias jets (minijets) appear as both angular correlations on $(\eta_\Delta,\phi_\Delta)$ and momentum correlations on $(p_t,p_t)$. The pQCD expectation for jet angular correlations is a nominally symmetric same-side peak centered at the angular origin (intra-jet correlations) and a broad (on azimuth) away-side 1D ridge (inter-jet correlations from back-to-back jets) peaked at $\phi_\Delta = \pi$. Jet correlations on  $(p_t,p_t)$ (hard component) from \pp collisions are observed to be peaked near \pt = 1 GeV/c and well resolved from a soft component which does not extend above 0.5 GeV/c~\cite{porter2,porter3,pythia,hijing}. 
In \pp and peripheral \auau collisions the minijet hypothesis describes all data well. In more-central \auau collisions jet-like angular correlations are modified~\cite{axialci,daugherity}, but the jet hypothesis is certainly not excluded by the data.

In the context of hydro models the elongated SS 2D peak (``soft ridge'') has been attributed to radial flow coupled to conjectured elongated structure in the initial-state transverse configuration space. However, there is presently no evidence to compel preference of flow conjectures over a jet mechanism and substantial evidence that disfavors them~\cite{glasma,glasma2}.

\subsection{Jet-related terminology and partitions}

Misleading terminology may produce confusion regarding jets. Problems arise from arbitrary partition of kinematic variables $p_t$ and $\eta$ based on conjectures about jet properties. Reference is made to ``high-\pt jets'' or ``semihard fragments.'' Hadrons are described as ``soft'' or ``hard'' based on partition of the $p_t$ axis. Low-\pt (``soft'') hadrons (e.g., below 2 GeV/c) are excluded from jets, reserved instead for ``bulk'' and ``flow'' phenomena. But for jets of any energy the majority of hadron fragments fall below 2 GeV/c~\cite{eeprd,fragevo}. The terms soft and hard should describe IS momentum transfers between projectiles (hadrons or partons), not FS hadron fragments.

The pair $\eta$ acceptance is arbitrarily divided into ``short-range'' and ``long-range'' regions, the former reserved for jet interpretations, the latter associated with conjectured bulk collective phenomena based on questionable causality arguments. Long-range correlations are attributed to ``early-time'' interactions and are then said to be a manifestation of collective motion best described by the lowest few azimuth multipoles or ``harmonic flows.'' But $\eta$ is a measure of polar angle, not momentum per se. Jet structure at large $\eta$ is not precluded.

Biased terminology and unjustified partitions may incorrectly consign jets to a small fraction of the final state, nominally within $|\eta_\Delta| < 1$ (or less) and $p_t > 2$ GeV/c (or greater). The SS peak in more-central \aa collisions is arbitrarily partitioned into jet-like and ridge-like components with distinct production mechanisms. Any deviation of jet morphology from a notional ideal is attributed to nonjet (flow) mechanisms. Projection of some fraction or all of the SS peak structure onto 1D azimuth leads to (mis)attribution of jet correlations to flows via Fourier analysis.

\subsection{Minimum-bias vs $p_t$-conditional jets} 

Factorization of two-particle momentum space into 2D subspaces $(p_t,p_t)$ or $(y_t,y_t)$ (transverse momentum or rapidity correlations) and $(\eta_\Delta,\phi_\Delta)$ (angular correlations) retains almost all correlation information~\cite{porter2,porter3}. Minimum-bias angular correlations correspond to an integral over the entire $(p_t,p_t)$ space. So-called ``trigger-associated'' correlations correspond to cut conditions (typically asymmetric) imposed on $(p_t,p_t)$.

Isolation of nominal jet structure from nonjet structure and combinatorial background typically follows different strategies in the two cases. 
In minimum-bias analysis a sibling-pair distribution containing correlations of interest is compared to a mixed-pair reference which is nominally uncorrelated. The net correlation structure is analyzed by 2D model fits which establish a zero offset from the fit, as described in Sec.~\ref{angcorrdata}. For $p_t$-conditional or trigger-associated analysis the sibling-pair distribution is typically projected onto 1D azimuth difference. A ZYAM background, estimated by the ZYAM offset criterion and published $v_2$ data, is subtracted from the sibling-pair distribution.
The two analysis methods may lead to very different inferred correlation structure and interpretations.

It is possible to apply the sibling-mixed model-fit procedure to conditional angular correlations based on binning $(p_t,p_t)$ or $(y_t,y_t)$ as well. In that case the results may be very different from what is inferred by ZYAM subtraction, even when the $p_t$ cuts are identical. In particular, jet yields may be much larger in the former case and the correlation shapes much different~\cite{tzyam}.

\subsection{Relating the SS 2D peak to azimuth multipoles}

We have demonstrated that all azimuth multipoles can be predicted by measured properties of the SS 2D peak, AS dipole and nonjet quadrupole.  The systematic variation of just five model parameters controls all 1D multipole phenomenology in nuclear collisions, whatever the physical interpretation. ``Higher harmonic flows'' (triangular flow, etc.) are determined completely by the three-parameter SS 2D peak. The 2D peak and the claimed flow system are therefore equivalent {\em within the 1D projection}. However, the three-parameter SS 2D peak (combined with nonjet quadrupole and AS dipole) is a complete representation of all 1D multipoles, whereas the 1D multipole system {\em cannot reconstruct} the SS 2D peak and is therefore incomplete. 
Based on full 2D correlations we conclude that all structure assigned in 1D to ``higher harmonics'' has a strong 2D dependence (large curvature on $\eta_\Delta$) not expected from flow mechanisms (Sec.~\ref{2dquad}).

 \section{Summary}

Two notable features have been reported in 2D [on $(\eta,\phi)$] angular correlations at RHIC: a same-side (SS) 2D peak elongated on pseudorapidity $\eta$ in more-central \auau collisions and an away-side (AS) 1D peak on $\phi$ seemingly distorted after subtraction of a combinatoric background. The apparent AS distortion appears as a double peak near $\pi$, interpreted by some as evidence for Mach cone formation in a dense medium. 

Recently, explanations for both apparent distortions have been proposed based on conjectured initial-state geometry multipoles combined with collective expansion to produce a long-range (on $\eta$) final-state quadrupole (increased elliptic flow) and sextupole (triangular flow) in the final state. Initial-state multipoles are said to represent fluctuations in the \aa overlap geometry; final-state multipoles are described as higher harmonic flows.

In this article we introduce the measured systematics of the SS 2D peak, AS dipole and nonjet azimuth quadrupole on \aa centrality. We show that the properties of those three structures predict all azimuth multipoles that can be derived from Fourier analysis of the 1D projection onto azimuth of 2D angular correlations. We show that some analysis methods relying on 1D projection can confuse Fourier components of a 1D Gaussian with nominally independent other elements (e.g., flow and nonflow).

We have reviewed the recent introduction of triangular flow and higher harmonic flows as a proposed explanation for elongation of the SS 2D peak. We show that the measured 2D peak properties do predict all triangular flow data $v_3$ and higher harmonics $v_4$, etc., both at 200 GeV, and with minor modification at LHC energies. We also demonstrate that Fourier analysis alone cannot describe 2D angular correlations. The 1D Fourier model is falsified by 2D data.  The 2D peak itself is then a more efficient {\em and complete} representation of 2D angular correlations than the separate azimuth multipoles derived from Fourier analysis of the 1D projection.

The surprising abundance of minijets in FS \aa correlation structure has presented a serious problem for proponents of hydro-dominated RHIC collisions. Observed minijet systematics are consistent with negligible loss of minimum-bias scattered partons to thermalization even in central \auau collisions, leading to a recent trend to reinterpret nominal jet correlations as flow manifestations. Based on Fourier analysis of 1D projections onto azimuth collisions are said to be flow-dominated,  with minor additional contributions characterized as ``nonflow.'' We have demonstrated that 1D projection and Fourier coefficients alone provide an incomplete data description susceptible to incorrect inferences.

Any comprehensive description of high-energy nuclear collisions must accommodate {\em two-dimensional} pQCD jet structure in \pp and peripheral \aa collisions and admit the possibility of modified jet formation in more-central \aa collisions. The full structure of the same-side 2D peak on $\eta$ should be acknowledged as distinct from other correlation components. The centrality dependence of individual correlation components must be considered. The particle yield associated with the same-side peak (nominally jet fragments predicted by pQCD) should be described quantitatively. The same-side peak does have well-defined Fourier components when projected onto 1D azimuth, but interpretation of any single Fourier term as representing flow can be questioned.

This work was supported in part by the Office of Science of the U.S. DOE under grant DE-FG03-97ER41020.

\begin{appendix}

 \section{correlation models} \label{modeling}

2D angular correlations are defined on difference-variable space $(\eta_\Delta,\phi_\Delta)$ within angular acceptance $(\Delta \eta,\Delta \phi)$. Accurate $\eta_\Delta$ dependence is essential to distinguish among data model elements and hadron production models.  1D projections abandon key information and greatly reduce the ability of projected data to test theoretical models. Any structure projected onto periodic 1D \dphi can be represented by a Fourier series, but interpretation of the coefficients may be ambiguous. 

\subsection{Model functions and 2D histograms}

2D correlation structure within some limited $\eta$ acceptance can be separated into $\eta$-independent and $\eta$-dependent components, as in Eq.~(\ref{estructfit}). The SS 2D peak centered at the origin always has a large curvature on $\eta_\Delta$ well-described by a Gaussian. The two sinusoids (AS dipole and nonjet quadrupole) have negligible curvature on $\eta_\Delta$ within the STAR TPC acceptance. The strong curvature on \deta of the SS 2D peak contradicts any description relying solely on azimuth multipoles uniform on $\eta_\Delta$.  Each term in Eq.~(\ref{estructfit}) is associated with a corresponding correlation model component on transverse momentum $(p_t,p_t)$, which may help to reveal the source mechanism and falsify incorrect models.

In elementary high-energy hadron collisions we expect significant jet structure, including a narrow SS 2D peak at the origin and an AS 1D peak (ridge) at \dphi = $\pi$ representing momentum conservation for IS pairs of large-angle-scattered partons~\cite{ua1,axialci}. The 2D data model which arose from empirical study of angular correlations in \pp and peripheral \aa collisions  includes three major elements. The minimum-bias jet contribution accounts for two terms in Eq.~(\ref{histo})~\cite{porter2,porter3}. In more-central  \aa collisions  an additional {\em nonjet} azimuth quadrupole term independent of $\eta_\Delta$ emerges as a third term.  

\subsection{Optimum data model}

Is there a unique 2D data model? If not, can a ``best'' of several models be identified? According to Ockham's razor the smallest number of model parameters is preferred. All model elements should be {\em necessary} and the overall model should be {\em sufficient}. Necessity of a term in Eq.~(\ref{histo}) can be tested by omitting it individually from the fit model. If the corresponding shape appears in the fit residuals the term is {\em necessary} to the data. If a combination of necessary terms completely describes the data, with no significant structure in the fit residuals, that fit model is {\em sufficient}. A necessary and sufficient (\mbox{N-S}) model is statistically equivalent to the 2D data and may then be the ``best'' (possibly unique) model. Uniqueness may be challenged, but competing models must describe {\em all aspects of the 2D data}, including $p_t$ dependence, centrality dependence and collision-energy dependence, {\em as well as or better than} the N-S model that is challenged.

\subsection{1D projections and Fourier series}

2D angular correlations can be projected onto 1D \deta or $\phi_\Delta$.  
The projection onto \dphi is conventionally denoted dihadron correlations with ZYAM subtraction~\cite{tzyam}. 
1D projection is also implicit in nongraphical numerical methods (conventional $v_2$ analysis)~\cite{2004}. 
Any 2D data projected onto periodic 1D azimuth can be represented by a single Fourier series, but such 1D projections may superpose multiple 2D model elements onto each Fourier coefficient (e.g., flow and nonflow) leading to confused interpretations.

Replacing the SS 2D Gaussian by a 1D Fourier series increases the number of model parameters and degrades the 2D data description, and the 1D Fourier series  cannot describe the unprojected 2D angular correlations. Adding additional Fourier terms to the N-S model of Eq.~(\ref{estructfit}) would
 degrade the capacity of the model to test hypotheses. More than one model element may then represent the same data feature leading to redundancy.  Large excursions of fitted parameters may give the misleading impression that the original N-S model is an insufficient data description.

\section{Flow and Nonflow} \label{nonflowapp}

The terms ``flow'' and ``nonflow'' present an important definition problem for angular correlations at the RHIC. Flow in  that context refers to ``elliptic flow,'' a conjectured hydrodynamic phenomenon~\cite{2004}. Conventional descriptions of elliptic flow assume that projected azimuth correlations are well described by a total azimuth quadrupole $v_2 \propto \cos(2\phi_\Delta)$ whose dominant contribution is elliptic flow. Nonflow is assumed to be a sum of small contributions to the total quadrupole from extraneous structures (e.g., resonance correlations and possibly jets) which contribute small biases to $v_2$ measurements.

Definition problems arise from a breakdown of basic assumptions. Minimum-bias angular correlations are actually dominated by a jet-like structure clearly apparent in 2D correlations which has a large curvature on $\eta_\Delta$ not expected for hydro phenomena. The jet-like structure projected onto azimuth has a substantial quadrupole component in its Fourier representation. A second quadrupole component with negligible curvature on $\eta_\Delta$ distinct from jet-like structure is also apparent. The two quadrupole contributions may be better described as jet-related and nonjet quadrupoles. However, they also correspond systematically to the terms nonflow and flow respectively.

Several methods have been invoked in attempts to reduce nonflow contributions to measured $v_2$. Higher cumulants such as $v_2\{4\}$ were expected to reduce sensitivity to few-particle correlations (such as jets) but may also be biased (negatively) by conjectured flow fluctuations.  Combinations of $v_2$ measurements have been used in attempts to estimate nonflow bias~\cite{2004}. A favored method relies on $\eta$ exclusion cuts. An exclusion zone $|\eta_\Delta| < O(1)$ is expected to eliminate nonflow given the assumption that nonflow is always a ``short-range'' phenomenon. With the appearance of the ridge in triggered dihadron correlations and $\eta$ elongation of the SS 2D peak in minimum-bias correlations the possibility of jet structure (nonflow) outside the $\eta$ exclusion zone has emerged.

In a recent paradigm shift {\em maximum} possible $v_m$ values are sought using $v_m\{2\}$ and higher multipoles are included:
``This [extension to higher $m$] completely removes any assumption about the nature of non-flow correlations, except that [sic] is negligible at large relative pseudorapidity. As we have seen, this is a well-supported assumption''~\cite{luzum}. That strategy marks a reversal of previous attempts to discriminate ``elliptic flow'' (nonjet quadruple) from ``nonflow'' (jet-related quadrupole). It is assumed that any deviation from jet structure observed in \pp collisions, or an idealized (notional) jet definition, is a manifestation of flows.

 \section{2D multipole correlations} \label{2dmult}

We consider the 2D consequences of manipulating 1D azimuth multipoles. Simulations are based on measured minimum-bias angular correlations from central \auau collisions~\cite{axialci,daugherity}. Curvature effects are emphasized by extrapolating the data parametrization to pair acceptance $\eta_\Delta \in [-4,4]$.  The STAR TPC acceptance extends to $\eta_\Delta \in [-3,3]$ but is typically limited to [-2,2] for correlation analysis to reduce small artifacts at the acceptance edges. 

 \subsection{Multipole decomposition of the SS 2D peak}

\begin{figure}[h]
 \includegraphics[width=1.6in,height=1.5in]{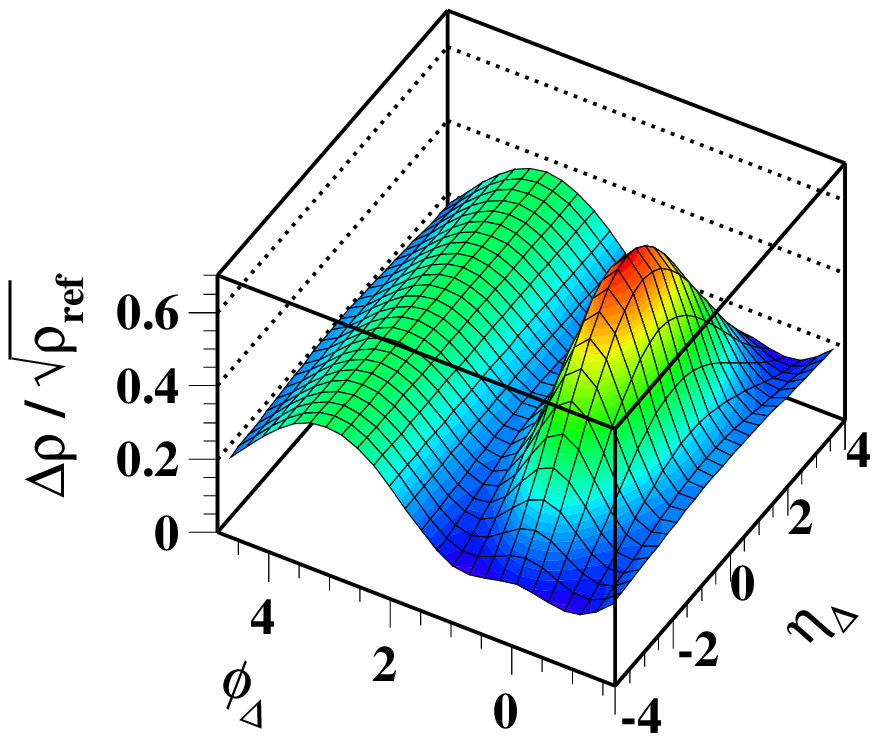}
  \includegraphics[width=1.6in,height=1.5in]{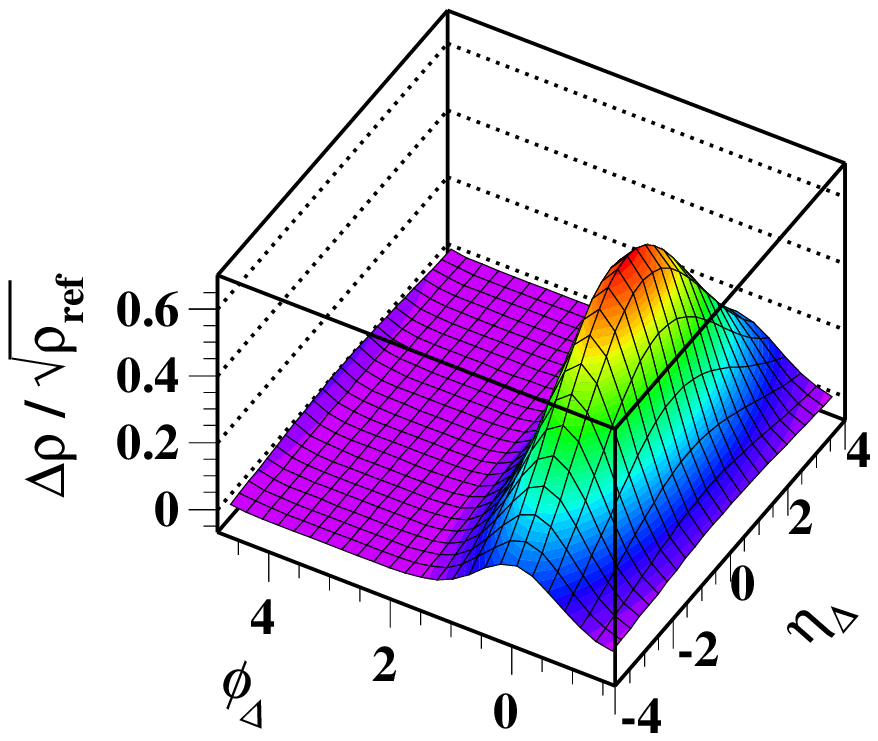}
\caption{\label{jetpeaks1} (Color online)
Left: Simulated angular correlations from 200 GeV \auau collisions with centrality $\nu = 6$ ($b = 0$) extrapolated to $\eta \in [-2,2]$ with nonjet components (quadrupole, 1D Gaussian on $\eta_\Delta$) subtracted to reveal jet correlations.
Right: The same histogram with the away-side dipole term subtracted to isolate the same-side 2D jet peak.
 } 
 \end{figure}

Figure~\ref{jetpeaks1} (left panel) shows simulated data (nominal jet angular correlations) from central 200 GeV \auau collisions ($\nu = 6$, panel (d) in Fig.~\ref{2dcorr}). The nonjet quadrupole component $A_Q\{2D \}$ for that centrality is essentially zero. There are then two correlation components, the SS 2D peak and the AS dipole.
Figure~\ref{jetpeaks1} (right panel) shows the left panel with the AS dipole removed to isolate the SS 2D peak.

\begin{figure}[h]
  \includegraphics[width=1.65in,height=1.6in]{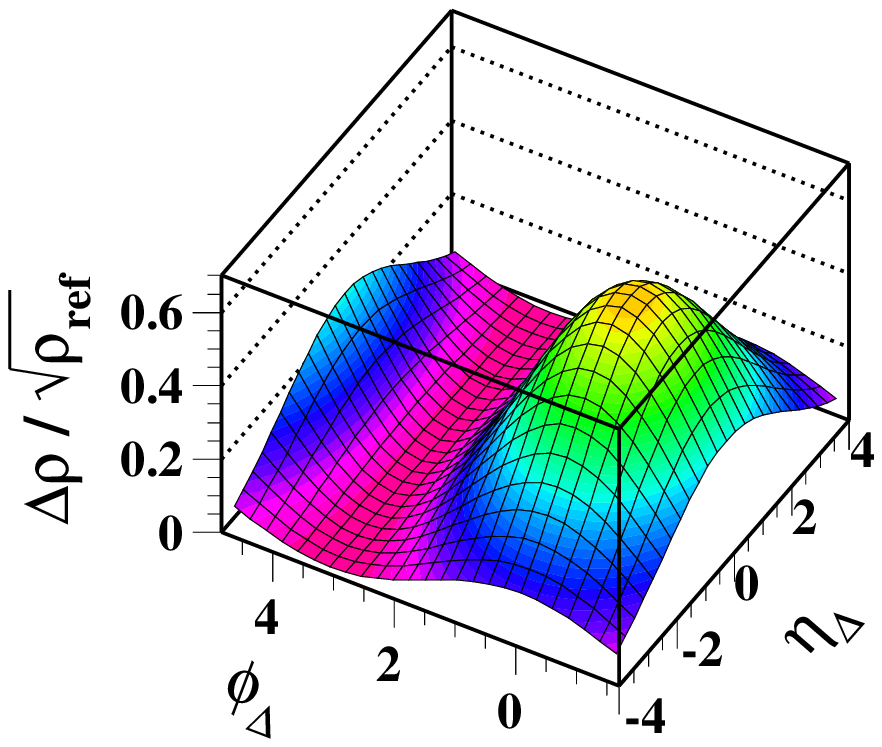}
\put(-110,100){$\bf A_D\{SS\}$}
  \includegraphics[width=1.65in,height=1.6in]{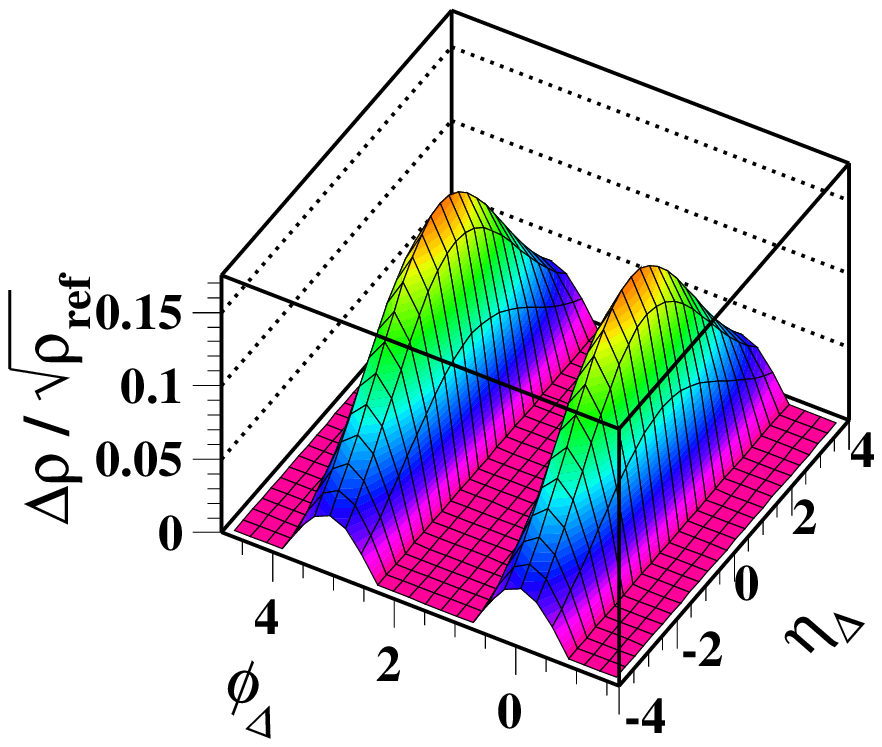}
\put(-110,100){$\bf A_Q\{SS\}$}\\
 \includegraphics[width=1.65in,height=1.6in]{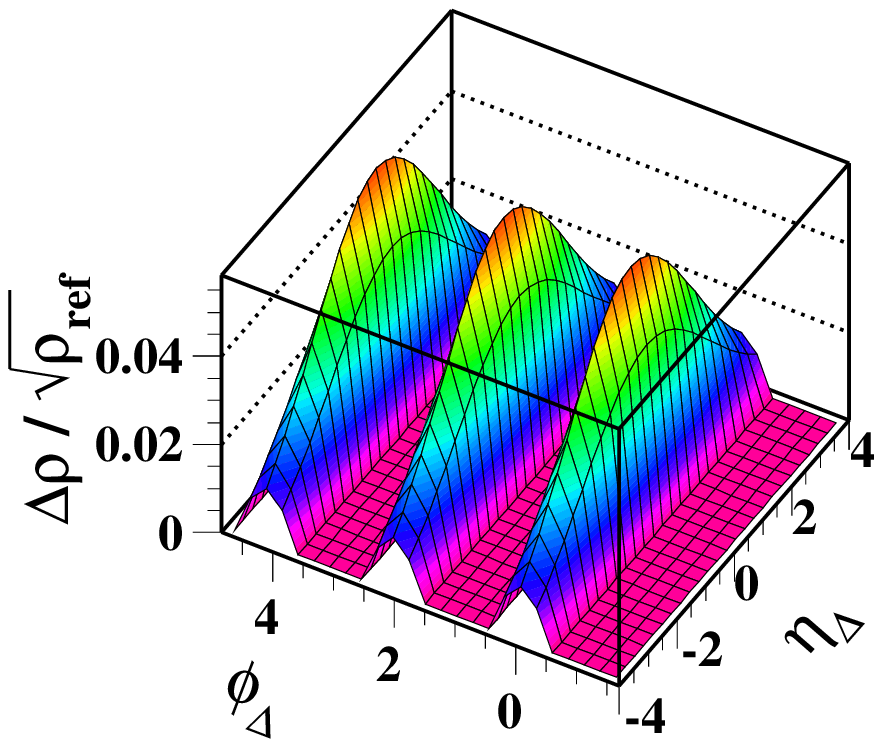}
\put(-110,100){$\bf A_S\{SS\}$}
  \includegraphics[width=1.65in,height=1.6in]{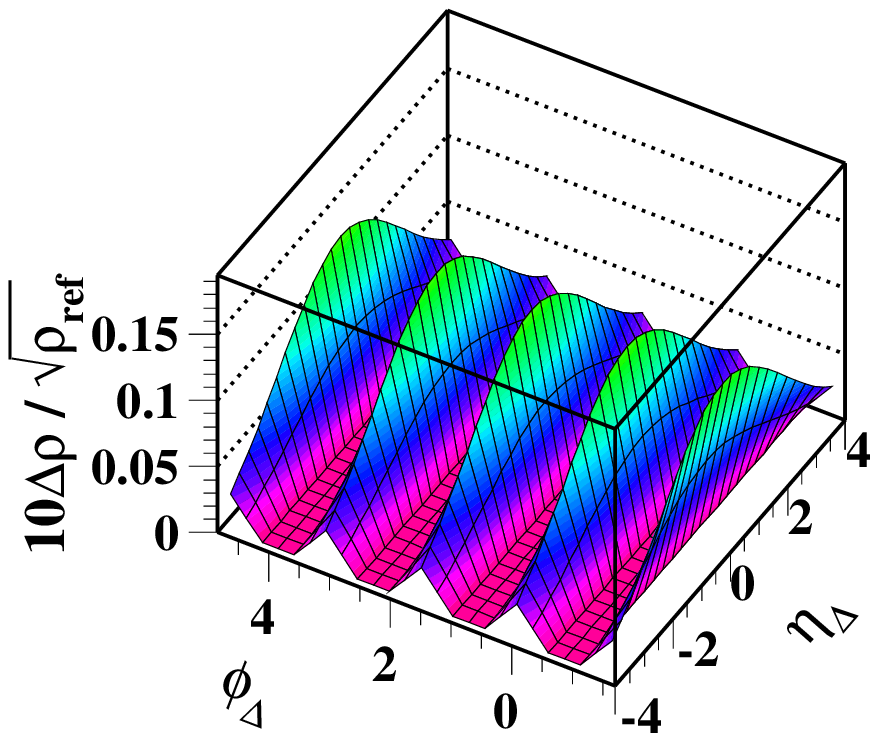}
\put(-110,100){$\bf A_O\{SS\}$}
\caption{\label{mults1} (Color online)
Same-side 2D multipole components $A_X\{SS\}$ of the SS 2D peak in Fig.~\ref{jetpeaks1} (right panel): $X \rightarrow$  dipole D, quadrupole Q, sextupole S and octupole O.
 } 
 \end{figure}

Arguments in Refs.~\cite{gunther},~\cite{luzum} and others actually relate to the SS 2D peak, as we have shown in Sec.~\ref{predict}. The SS peak for minimum-bias angular correlations is consistent with a single 2D Gaussian in more-central \auau collisions. The SS peak retains a large curvature on $\eta_\Delta$ for all collision conditions. Figure~\ref{mults1}  shows four multipoles (D,Q,S,O) from a 2D Fourier decomposition of the SS peak. Those structures correspond to the 1D projections in Fig.~\ref{fourier} (right panel, dotted curves for $m=1...4$).

\subsection{2D ZYAM subtraction}

\begin{figure}[h]
   \includegraphics[width=1.65in,height=1.65in]{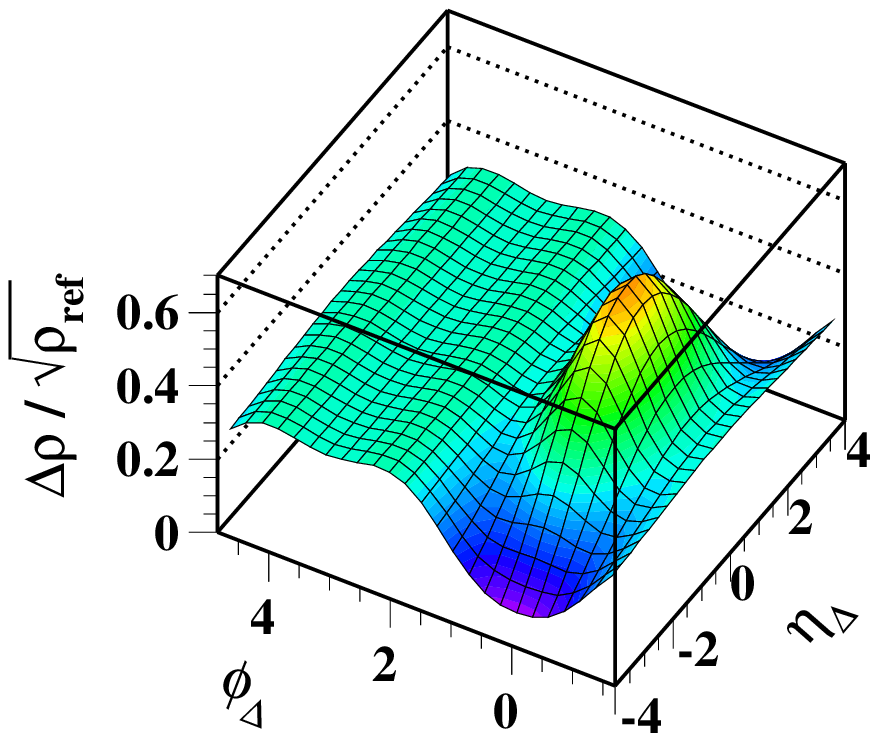}
   \includegraphics[width=1.65in,height=1.65in]{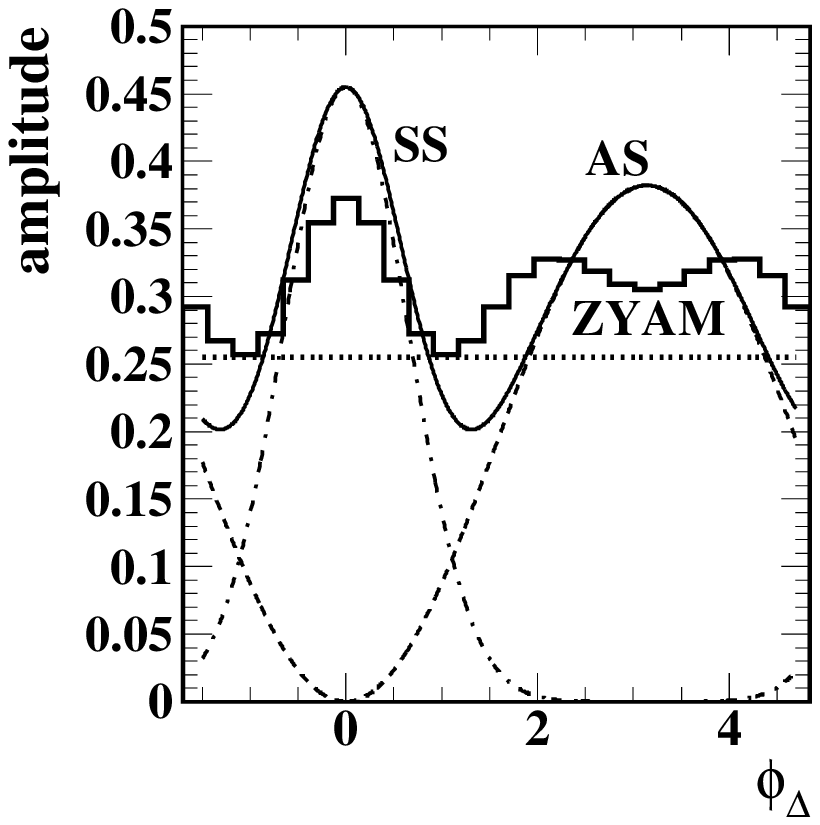}
\caption{\label{zyam3} (Color online)
Illustration of 2D ZYAM subtraction with 1D quadrupole (75\% of 1D $v_2^2\{SS\}$ quadrupole amplitude)
Left: 2D angular correlations resulting from ZYAM subtraction applied to the data in Fig.~\ref{jetpeaks1} (left panel),
Right: 1D projection of the histogram in the left panel (bold solid histogram), the actual SS peak (dash-dotted curve) and AS peak (dashed curve) with their sum (light solid curve).
 } 
 \end{figure}

We can use the 2D multipoles to illustrate the effect on 2D angular correlations of 1D ZYAM subtraction. Figure~\ref{zyam3} (left panel) shows 2D correlations from Fig.~\ref{jetpeaks1} (left panel) after subtraction of 75\% of the SS 2D peak quadrupole $V_2\{SS\}$, typical $v_2$ oversubtraction from nonflow bias in ZYAM subtraction. The AS dipole is somewhat distorted, but the SS 2D peak is still prominent.

 Figure~\ref{zyam3} (right panel) shows the 1D projection of the left panel as the bold histogram labeled ZYAM. The baseline estimate for those data according to the ZYAM principle (the ``background'' to be subtracted) is denoted by the dotted line. The actual jet structure is indicated by the dash-dotted and dashed curves for SS and AS peaks respectively relative to true zero. The structure of the ZYAM histogram is typical of what emerges from such analysis~\cite{tzyam}. The apparent magnitude of the jet yield is greatly underestimated as a result of the ZYAM analysis but is still clearly evident in 2D angular correlations.

 \begin{figure}[h]
   \includegraphics[width=1.65in,height=1.6in]{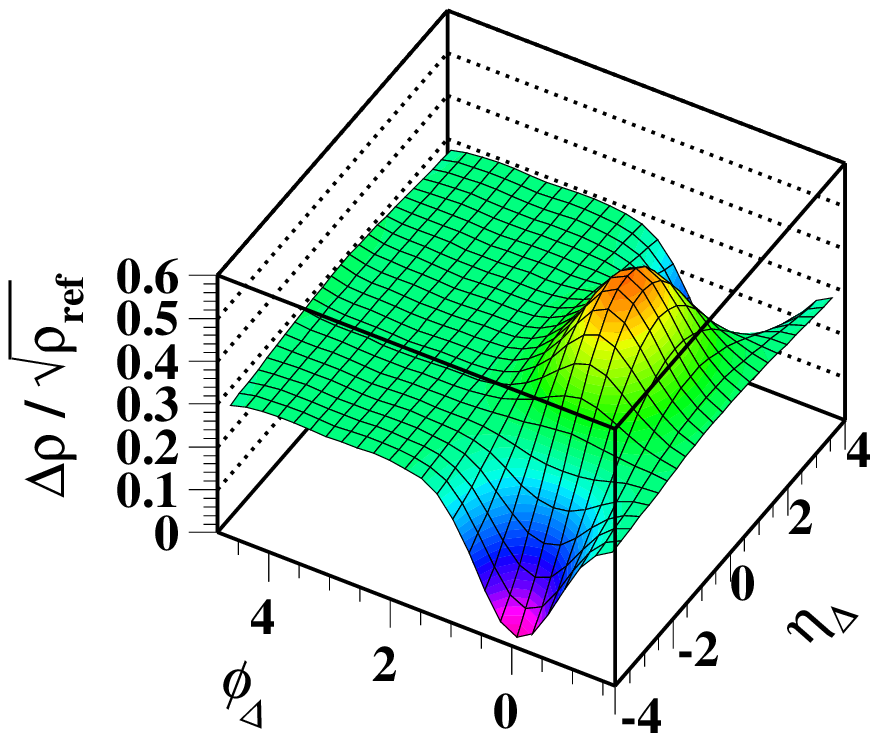}
 \includegraphics[width=1.65in,height=1.6in]{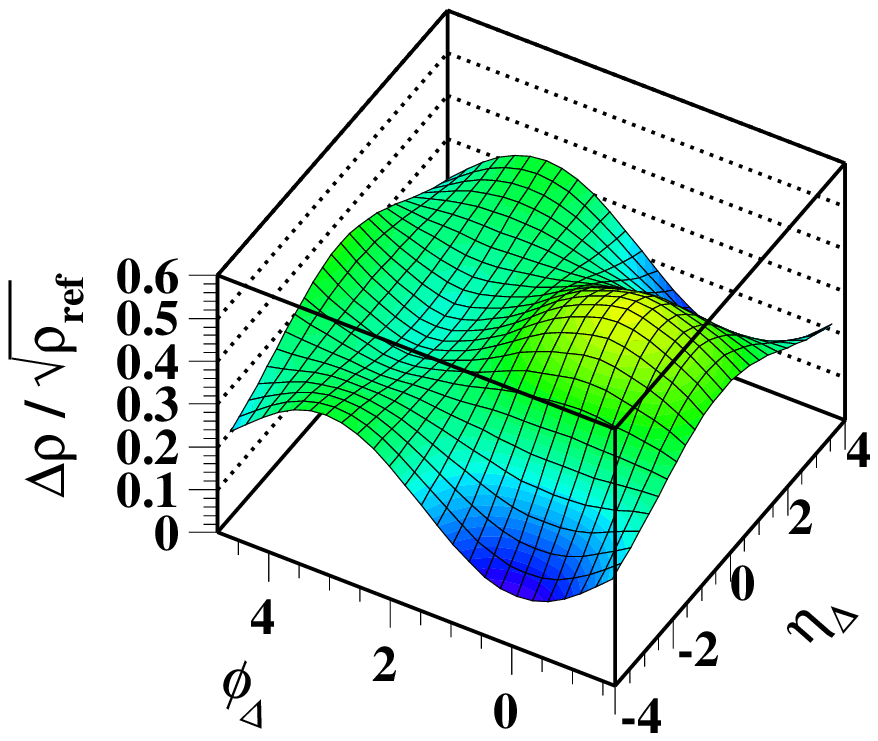}
\caption{\label{zyam4} (Color online)
2D histograms with 1D or 2D multipoles for $m = 2...4$ subtracted. 
Left: 2D histogram in Fig.~\ref{jetpeaks1} after subtracting three SS 1D multipoles. The AS dipole is canceled and a negative SS 1D peak is superposed on the positive SS 2D peak.
Right: The same 2D histogram after subtracting three SS 2D multipoles. The SS 2D peak is canceled and a positive SS 2D dipole is superposed on the AS 1D dipole. Both histograms project to flat 1D histograms on $\phi_\Delta$.
 } 
 \end{figure}

In Fig.~\ref{zyam4} we simulate the suggestion in Ref.~\cite{luzum} to  extend the ZYAM procedure by subtracting multipoles for $m=2,3,4$. In the left panel we subtract 1D multipoles, in the right panel the 2D multipoles from Fig.~\ref{mults1}.
The three SS multipoles combined are approximately equivalent to the SS peak minus the SS dipole, for either the 1D or 2D case. 
  As a consequence of 1D subtraction (left panel) the AS dipole is canceled by the SS dipole and the projected SS 1D Gaussian, extrapolated across the $\eta$ acceptance, is subtracted from the SS 2D peak. The highly-structured 2D histogram, when projected onto 1D azimuth, is flat (uniform on azimuth) with value $\approx 0.3$.

As a consequence of 2D subtraction (right panel) the SS 2D peak is canceled and the SS 2D dipole is added to the AS 1D dipole.  The highly-structured 2D histogram, when projected onto azimuth, is again fat with value $\approx 0.3$. For this demonstration the AS dipole amplitude $A_D$ was adjusted to just cancel the projected SS 2D peak dipole ($A_D \rightarrow F_1 G A_{2D}$), resulting in flat 1D projections after subtraction. In real systems the AS dipole amplitude tends to be somewhat larger relative to the SS peak. The same analysis would then result in a small net AS dipole, interpreted by some as ``global momentum conservation.'' By subtracting the three multipoles all evidence of jet structure can be removed from projected 1D azimuth correlations. But the jet structure is still clearly evident in 2D correlations, albeit strongly distorted.

\section{$\bf \eta$ exclusion cuts} \label{etacuts}

We seek an expression for factor $G(\sigma_{\eta_\Delta};\eta_1,\eta_2)$ relating 
the amplitude of a 2D Gaussian projected onto $\phi_\Delta$ with $\eta$ exclusion cuts denoted by $(\eta_1,\eta_2)$ imposed to the amplitude of its 1D Gaussian projection. The cut system consists of accepted $\eta$ intervals $[-\eta_2,-\eta_1]$ and $[\eta_1,\eta_2]$ symmetric about the origin. Defining $\Delta \eta = \eta_2 - \eta_1$ and  $s = \sqrt{2} \sigma_{\eta_\Delta}$ the projection factor is given by
\bea
G(\sigma_{\eta_\Delta};\eta_1,\eta_2)  &=&  \frac{A + B}{C + D},
\eea
with
\bea
A &=& \int_{2\eta_1}^{\eta_1+\eta_2} dx (x-2\eta_1) e^{-x^2 / s^2} \\ \nonumber
B &=& \int_{\eta_1 + \eta_2}^{2\eta_2} dx (2\eta_2 - x) e^{-x^2 / s^2} \\ \nonumber
C &=& \int_{2\eta_1}^{\eta_1+\eta_2} dx (x-2\eta_1)  \\ \nonumber
D &=& \int_{\eta_1 + \eta_2}^{2\eta_2} dx (2\eta_2 - x)   
\eea
Intermediate expressions are
\bea
E(a,b) &=& \int_a^b dx e^{-x^2 / s^2} \\ \nonumber
&=& \frac{\sqrt{\pi} s}{2} \left[ \erf (b / s) - \erf(a/s) \right] \nonumber \\ \nonumber
F(a,b) &=&  \int_a^b dx x e^{-x^2 / s^2} \\ \nonumber
&=&\frac{s^2}{2} \left[  e^{-a^2 / s^2} - e^{-b^2 / s^2}  \right].
\eea
With $C+D = \Delta \eta^2$ the other terms are given by
\bea
A &=& F(2\eta_1,\eta_1 + \eta_2) - 2\eta_1 E(2\eta_1,\eta_1 + \eta_2)\\ \nonumber
B &=& 2\eta_2 E(\eta_1 + \eta_2,2\eta_2) - F(\eta_1 + \eta_2,2\eta_2),
\eea
 and $G(\sigma_{\eta_\Delta};\eta_1,\eta_2)$ is determined. The limiting case $\eta_1 \rightarrow -\eta_2$ corresponds to $G(\sigma_{\eta_\Delta};\eta_1,\eta_2) \rightarrow G(\sigma_{\eta_\Delta};\Delta\eta)$ in Eq.~(\ref{gfac}).

\end{appendix}


\end{document}